\documentclass[aps,showpacs,twocolumn,twoside,groupedaddress]{revtex4}
\usepackage{graphicx}
\usepackage{epsfig}
\usepackage{epsf}
\usepackage{amssymb}
\usepackage{epstopdf}
\usepackage{amsmath}
\usepackage{braket}
\usepackage{multirow}
\usepackage{array}
\usepackage{mathrsfs}
\usepackage{bm}
\let\vec\bm

\makeatletter
\renewcommand{\fnum@figure}{Fig. \thefigure}
\makeatother

\begin{document}

\title{Waveguide quantum electrodynamics in squeezed vacuum}
\author{Jieyu You, Zeyang Liao\footnote{zeyangliao@physics.tamu.edu}, Sheng-Wen Li, and M. Suhail Zubairy\footnote{zubairy@physics.tamu.edu}}
\affiliation{Institute for Quantum Science and Engineering (IQSE) and Department of Physics and Astronomy, Texas A$\&$M University, College Station, TX 77843-4242, USA }

\begin{abstract}

We study the dynamics of a general multi-emitter system coupled to the squeezed vacuum reservoir and derive a master equation for this system based on the Weisskopf-Wigner approximation. In this theory, we include the effect of positions of the squeezing sources which is usually neglected in the previous studies. We apply this theory to a quasi-one-dimensional waveguide case where the squeezing in one dimension is experimentally achievable. We show that while dipole-dipole interaction induced by ordinary vacuum depends on the emitter separation, the two-photon process due to the squeezed vacuum depends on the positions of the emitters with respect to the squeezing sources. The dephasing rate, decay rate and the resonance fluorescence of the waveguide-QED in the squeezed vacuum are controllable by changing the positions of emitters. Furthermore, we demonstrate that the stationary maximum entangled NOON state for identical emitters can be reached with arbitrary initial state when the center-of-mass position of the emitters satisfies certain condition. 

\end{abstract}
\maketitle 

\section{INTRODUCTION}

Due to the well known Purcell effect \cite{Purcell1946}, the spontaneous decay rate of an emitter can be modified by engineering the electromagnetic bath environment with which the emitters interact. One example of bath engineering is the squeezed vacuum. Although the squeezed vacuum does not change the density of the electromagnetic modes, it can still modify the decay rate of the emitter \cite{Gardiner1986, Collett1984, Zubairy1988}. A single emitter interacting with the squeezed vacuum has been widely studied \cite{Gardiner1987, Palma2, Palma1989}. However, there are only a few publications dealing with multiple emitters interacting with squeezed vacuum. Among these works, most are considering the case where emitters are separated by much less than an optical wavelength which is the well known Dicke model \cite{Agarwal1990}. It is shown that in a broadband squeezed vacuum, emitter system evolves into a state whose properties are similar to those of the squeezed vacuum. Only a very few papers study the case when the separation between the emitters becomes important \cite{Ficek1990, Ficek1991, Goldstein1996}. It is found that the dipole-dipole interaction induced by ordinary vacuum depends on the relative emitter separation, while the interaction induced by the squeezed vacuum depends on the center of mass coordinate of the emitters. Since it depends on the position of the center of mass, the choice of the coordinate system should be no longer arbitrary. However, it is not yet clearly illustrated in these literature on how to choose the coordinate system. Actually, the dependence on the absolute position comes from the fact that the squeezed vacuum is not vacuum but generated by a coherent light source. The phase of a coherent source is important for the dynamics of the emitter system \cite{Das2008} and it is seldom considered in the previous literature. People usually thought this phase can be included in the phase of the correlation function. However, the phase in the correlation function is usually treated as a constant, while it can a function of position. In addition, the previous calculations mainly consider a broadband squeezing in all directions of the 3-dimensional (3D) space which is difficult to be experimentally realized.

Recently, photon transport in a one-dimensional (1D) waveguide coupled to quantum emitters (well known as ``waveguide-QED") has attracted much attention due to its possible applications in quantum device and quantum information \cite{Shen2005a, Shen2005b, Shen2007, Yudson2008, Zheng063816, Zhou2008, Shi205111, Chen2011, Liao2015, Shen2015, Liao2016b,  Liao2016, Roy2017}. In these previous studies, the photon modes in the waveguide are usually considered to be ordinary vacuum modes. The case when the waveguide modes are squeezed is seldom studied. In contrast to the 3D case, squeezing in 1D is more experimentally feasible. Suppression of the radiative decay of atomic coherence and the linewidth of the resonance fluorescence have been experimentally demonstrated in a 1D microwave transmission line coupled to single artificial atom \cite{Turchette1998, Kocabas2012, Murch2013, Toyli2016}. However, many-body interaction in a 1D waveguide-QED system coupled to squeezed vacuum has not yet been studied.

In this paper we consider the phase of the squeezing source and rederive the master equation for multi-atom dynamics in the squeezed vacuum based on the Weisskopf-Wigner approximation. We show that while the collective dipole-dipole interaction due to the ordinary vacuum depends on the emitter separation, the collective two-photon decay rate due to the squeezed vacuum largely depends on the center of mass position of the emitters relative to the squeezing source. We then apply this theory to the 1D waveguide-QED system with squeezing reservoir. Contrary to the traditional result that the dephasing rate of a single atom in the squeezed vacuum is a constant \cite{Zubairy1988, Scully1997}, our calculation shows that the dephasing rate is actually  position-dependent. As dipole-dipole interaction is involved, both emitter separation and center of mass coordinate can affect the decay rate, dephasing rate and the emitted resonance fluorescence spectrum. In addition, we also show that stationary quantum entanglement can be prepared in this system by the squeezing reservoir. The stationary maximum entangled NOON state can be approached if the center-of-mass of the emitters is at certain position.

This paper is organized as follows: In Sec. II, we introduce the Hamiltonian of the system and the modified mode function for the squeezed vacuum. In Sec. III, we derive the master equation for the emitter system in 3D case based on the Weisskopf-Wigner approximations. In Sec. IV, we consider the squeezing in a 1D waveguide-QED system where we show how the dephasing rate depends on the position of the atoms and we also show that stationary quantum entangled state can be prepared. Then, we analyze the properties of power spectrum under the effects of squeezed vacuum and dipole-dipole interaction. Finally, we summarize our results.

\section{Hamiltonian and mode function} 

We here consider $N_{a}$ identical two-level atoms located at $ \vec{r}_{i}$ ($i=1,\cdots,N_{a}$). Suppose that all the transition dipole moments $ \vec{\mu}_{i} $ have the same amplitude and direction. The atom-field system is described by the Hamiltonian \begin{equation}
  \label{eq1}
  \begin{gathered}
H=H_{A}+H_{F}+H_{AF}
 \end{gathered}
\end{equation}
where
$H_{A}=\sum_{i=1}^{N_a}\hbar\omega_i\left|e_{i}\right\rangle \left\langle e_{i}\right|$  
is the atomic Hamiltonian, and $ \left|e_{i}\right\rangle $ is the excited state of the $i$th atom with transition frequency $\omega_i$. Here, for simplicity, we assume that all the atoms have the same transition frequency, i.e., $\omega_{i}\equiv\omega_0$. The Hamiltonian of the EM field is
$H_{F}=\sum_{\vec{k}s}\hbar\omega_{\vec{k}s}(\hat{a}_{\vec{k}s}^{\dagger}\hat{a}_{\vec{k}s}+\frac{1}{2})$
where $\hat{a}_{\vec{k}s}$ and $\hat{a}_{\vec{k}s}^{\dagger}$ are the annihilation and creation operators of the filed mode with wavevector $ \vec{k}$, polarization $s$, and frequency $\omega_{\vec{k},s}$. The interaction Hamiltonian in electric-dipole approximation is
$H_{AF}=-i\hbar\sum_{\vec{k}s}\sum_{i=1}^{2}[\vec{\mu}_{i}\cdot\vec{u}_{\vec{k}s}(\vec{r}_{i})S_{i}^{+}\hat{a}_{\vec{k}s}+\vec{\mu}_{i}^{*}\cdot\vec{u}_{\vec{k}s}(\vec{r}_{i})S_{i}^{-}\hat{a}_{\vec{k}s}-H.c.]$
where $ \vec{\mu}_{i} $ is the electric dipole moment and $ S_{i}^{+} $ and $S_{i}^{-} $ are the raising and lowering operator for the $i$th atom. The mode function of the squeezed vacuum is given by 
\begin{equation}
  \label{eq2b}
  \begin{gathered}
\vec{u}_{\vec{k}s}(\vec{r}_{i})=\sqrt{\frac{\omega_{\vec{k}s}}{2\epsilon_{0}\hbar V}}\vec{e}_{ks}e^{i\vec{k}\cdot(\vec{r}_{i}-\vec{o}_{\vec{k}s})}
 \end{gathered}
\end{equation}
where $\vec{o}_{\vec{k}s} $ includes the effects of the initial phase and the position of the squeezing source with wavevector $ \vec{k} s$. Here we need to make two assumptions: first, one specific mode is generated from a single source, i.e., mode $ \vec{k}s$ is only generated from the source located at $\vec{o}_{\vec{k}s}$; second, the phases of all modes can be well defined by $ \vec{k}\cdot(\vec{r}-\vec{o}_{\vec{k}s}) $. In the ordinary vacuum or thermal reservoir, there is no source and we can set $\vec{o}_{\vec{k}s}=0$, so the mode function shown in Eq. \eqref{eq2b} is reduced to the normal cases \cite{Scully1997}. However, when the reservoir is produced by different sources with non-vanishing correlation function $\langle\hat{a}_{\vec{k}s}^{\dagger}\hat{a}_{\vec{k'}s'}^{\dagger}\rangle$ and $\langle\hat{a}_{\vec{k}s}\hat{a}_{\vec{k'}s'}\rangle$, for example, the squeezed vacuum reservoir, the spatial distribution of the source is important. Neglecting $\vec{o}_{\vec{k}s}$ in the mode function will lead to an ambiguity of physics where the emitters' coordiantes are not well defined \cite{Goldstein1996}. Therefore, the position of the source should be included in the mode function when the squeezed vacuum is considered. One can also add an additional global phase $e^{i\phi}$ to the mode function Eq. \eqref{eq2b}, but for simplicity\cite{Das2008}, we can set $\phi=0$.

\section{MASTER EQUATION}

In this section, we first derive the master equation of a multi-emitter system in a general 3D squeezed vacuum with the Hamiltonian shown in Eq. \eqref{eq1} and mode function shown in Eq. \eqref{eq2b}. The Hamiltonian in the interaction picture without rotating-wave approximation is given by
\begin{equation}
 \label{eq3}
 \begin{split}
V(t)=&-i\hbar\underset{\vec{k}s}{\sum}\underset{i}{\sum}[\vec{\mu}_{i}\cdot\vec{u}_{\vec{k}s}(\vec{r}_{i})S_{i}^{+}(t)\hat{a}_{\vec{k}s}(t) \\ &+\vec{\mu}_{i}^{*}\cdot\vec{u}_{\vec{k}s}(\vec{r}_{i})S_{i}^{-}(t)\hat{a}_{\vec{k}s}(t)-H.c.]
\end{split}
\end{equation}
where $ S_{i}^{\pm}(t)=S_{i}^{\pm}e^{\pm i\omega_{0}t} $, $ \hat{a}_{\vec{k}s}(t)=\hat{a}_{\vec{k}s}e^{-i\omega_{\vec{k}s}t} $, and $ \hat{a}_{\vec{k}s}^{\dagger}(t)=\hat{a}_{\vec{k}s}^{\dagger}e^{i\omega_{\vec{k}s}t} $. Different from Ref. \cite{Goldstein1996}, no rotating-wave approximation is made at this stage. The equation of motion for the reduced density matrix of the system is given by \cite{Scully1997} 
\begin{equation}
\label{eq4}
\begin{split}
\dot{\rho^{S}}=&-\frac{i}{\hbar}Tr_{R}[V(t),\rho^{S}(0)\otimes\rho^{F}(0)] \\ &-\frac{1}{\hbar^{2}}Tr_{R}\intop_{0}^{t}[V(t),[V(t-\tau),\rho^{S}(t-\tau)\otimes\rho^{F}(0)]]d\tau
\end{split}
\end{equation}
where $ \rho^{F} $ is the density matrix for the squeezed vacuum reservoir and is defined by $ \rho^{F}=\underset{\vec{k},s}{\prod}S_{\vec{k},s}\left|0_{\vec{k}_{0}\pm\vec{k}}\right\rangle \left\langle 0_{\vec{k}_{0}\pm\vec{k}}\right|S_{\vec{k},s}^{\dagger}$. The squeezed operator $S_{\vec{k},s}(\zeta)=\exp(\zeta^{*}a_{\vec{k}_{0}+\vec{k}}a_{\vec{k}_{0}-\vec{k}}-\zeta a_{\vec{k}_{0}+\vec{k}}^{\dagger}a_{\vec{k}_{0}-\vec{k}}^{\dagger})$ where $\zeta=re^{i\theta}$ is the squeezing parameter with the degree of squeezing $r$ and the squeezing phase $\theta$. For simplicity, we can also assume that $ck_{0}=\omega_0$, i.e.,  the center frequency of the squeezing field is equal to the transition frequency of the atom. 

For a squeezed vacuum reservoir, it can be shown that \cite{Scully1997}:

\begin{subequations}
\begin{align}
\left\langle a_{\vec{k},s}\right\rangle &=\left\langle a_{\vec{k},s}^{\dagger}\right\rangle =0 \label{eq51}\\
\left\langle a_{\vec{k},s}^{\dagger}a_{\vec{k}',s'}\right\rangle &=\sinh^{2}r\delta_{\vec{k}'\vec{k}}\delta_{ss'} \label{eq52}\\
\left\langle a_{\vec{k},s}a_{\vec{k}',s'}^{\dagger}\right\rangle &=\cosh^{2}r\delta_{\vec{k}'\vec{k}}\delta_{ss'} \label{eq53}\\
\left\langle a_{\vec{k},s}^{\dagger}a_{\vec{k}',s'}^{\dagger}\right\rangle &=-e^{-i\theta}\cosh(r)\sinh(r)\delta_{\vec{k}',2\vec{k}_{0}-\vec{k}}\delta_{ss'} \label{eq54}\\
\left\langle a_{\vec{k},s}a_{\vec{k}',s'}\right\rangle &=-e^{i\theta}\cosh(r)\sinh(r)\delta_{\vec{k}',2\vec{k}_{0}-\vec{k}}\delta_{ss'}   \label{eq55}
\end{align}
\end{subequations}
For simplicity, we can set the squeezing phase $\theta=0 $. On inserting these correlation functions into Eq. \eqref{eq4}, we can obtain the master equation (see Appendix A for the derivation):
\begin{widetext}
\begin{equation}
\label{eq6}
\begin{split}
\frac{d\rho^{S}}{dt}=&-i\underset{i\neq j}{\sum}\Lambda_{ij}[S_{i}^{+}S_{j}^{-},\rho^{S}]e^{i(\omega_{i}-\omega_{j})t}-\frac{1}{2}\underset{i,j}{\sum}\gamma{}_{ij}(1+N)(\rho^{S}S_{i}^{+}S_{j}^{-}+S_{i}^{+}S_{j}^{-}\rho^{S}-2S_{j}^{-}\rho^{S}S_{i}^{+})e^{i(\omega_{i}-\omega_{j})t} \\
&-\frac{1}{2}\underset{i,j}{\sum}\gamma{}_{ij}N(\rho^{S}S_{i}^{-}S_{j}^{+}+S_{i}^{-}S_{j}^{+}\rho^{S}-2S_{j}^{+}\rho^{S}S_{i}^{-})e^{-i(\omega_{i}-\omega_{j})t}-\frac{1}{2}\sum_{\alpha=\pm}\underset{i,j}{\sum}\gamma'_{ij}Me^{2\alpha ik_{0z}R}(\rho^{S}S_{i}^{\alpha}S_{j}^{\alpha}+S_{i}^{\alpha}S_{j}^{\alpha}\rho^{S}-2S_{j}^{\alpha}\rho^{S}S_{i}^{\alpha})
\end{split}
\end{equation}
\end{widetext}
where the first three terms are the same as in the thermal reservoir and the last term is the collective decay due to the squeezed vacuum. We have $M=\sinh(r)cosh(r)$ and average photon number $N=\sinh^{2}(r)$. The collective energy shifts $\Lambda_{ij}$ and decay rates $\gamma_{ij}$ due to the ordinary vacuum are given by \cite{Agarwal1974, Ficek2005}
\begin{align}
\Lambda_{ij}&=\frac{3}{4}\sqrt{\gamma_{i}\gamma_{j}}\{-(1-\cos^{2}\alpha)\frac{\cos(k_{0}r_{ij})}{k_{0}r_{ij}} \nonumber \\ & +(1-3\cos^{2}\alpha)[\frac{\sin(k_{0}r_{ij})}{(k_{0}r_{ij})^{2}}+\frac{\cos(k_{0}r_{ij})}{(k_{0}r_{ij})^{3}}]\}\\
\gamma_{ij}&=\sqrt{\gamma_{i}\gamma_{j}} F(k_{0}r_{ij})
\end{align}
where $\gamma=\frac{\omega_{0}^{3}\mu^{2}}{3\pi\epsilon_{0}\hbar c^{3}}$ is the spontaneous decay rate of the atom in ordinary vacuum and $F(x)=\frac{3}{2}\{(1-\cos^{2}\alpha)\frac{\sin x}{x}+(1-3\cos^{2}\alpha)[\frac{\cos x}{x^2}-\frac{\sin x}{x^3}]\}$. Different from the thermal reservoir terms, the squeezed vacuum can contribute to the additional collective two-photon decay rate of the system which is given by 
\begin{equation}
\label{ad9}
\gamma'_{ij}=\gamma e^{2ik_{0}R}F(k_{0} |\vec{r}_{i}+\vec{r}_{j}|).
\end{equation}
Thus, the collective decay due to the squeezed vacuum depends on the position of the center of mass of the emitters instead of their separation. One may think this reult is identical to the privious work\cite{Ficek1990,Ficek1991} except the phase $e^{2ik_{0}R}$, but that is not true. No matter how the coordinate system is built, to reach the neat form of Eq.\eqref{ad9}, $\vec{r}_i$ must still be interpreted as the displacement from the center of squeezing sources to the $i$th atom. 
 When their center of mass is at equal distances from all squeezing sources (i.e., $r_i+r_j=0$), the decay induced by the squeezing is the strongest due to the perfectly constructive interference of the two-photon excitation from all directions. It decreases when it deviates from the center due to the destructive interference. The master equation shown in Eq. (6) can be transformed to the Lindblad form \cite{Lindblad1976} and  the density matrix is positive definite which is proven in Appendix B. The phase factor $e^{2ik_{0z}R}$ can be effectively regarded as an controllable phase of $M$, which can be incorporated into $\theta$.

\section{Waveguide-QED in the squeezed vacuum}

\begin{figure}
\includegraphics[width=\columnwidth]{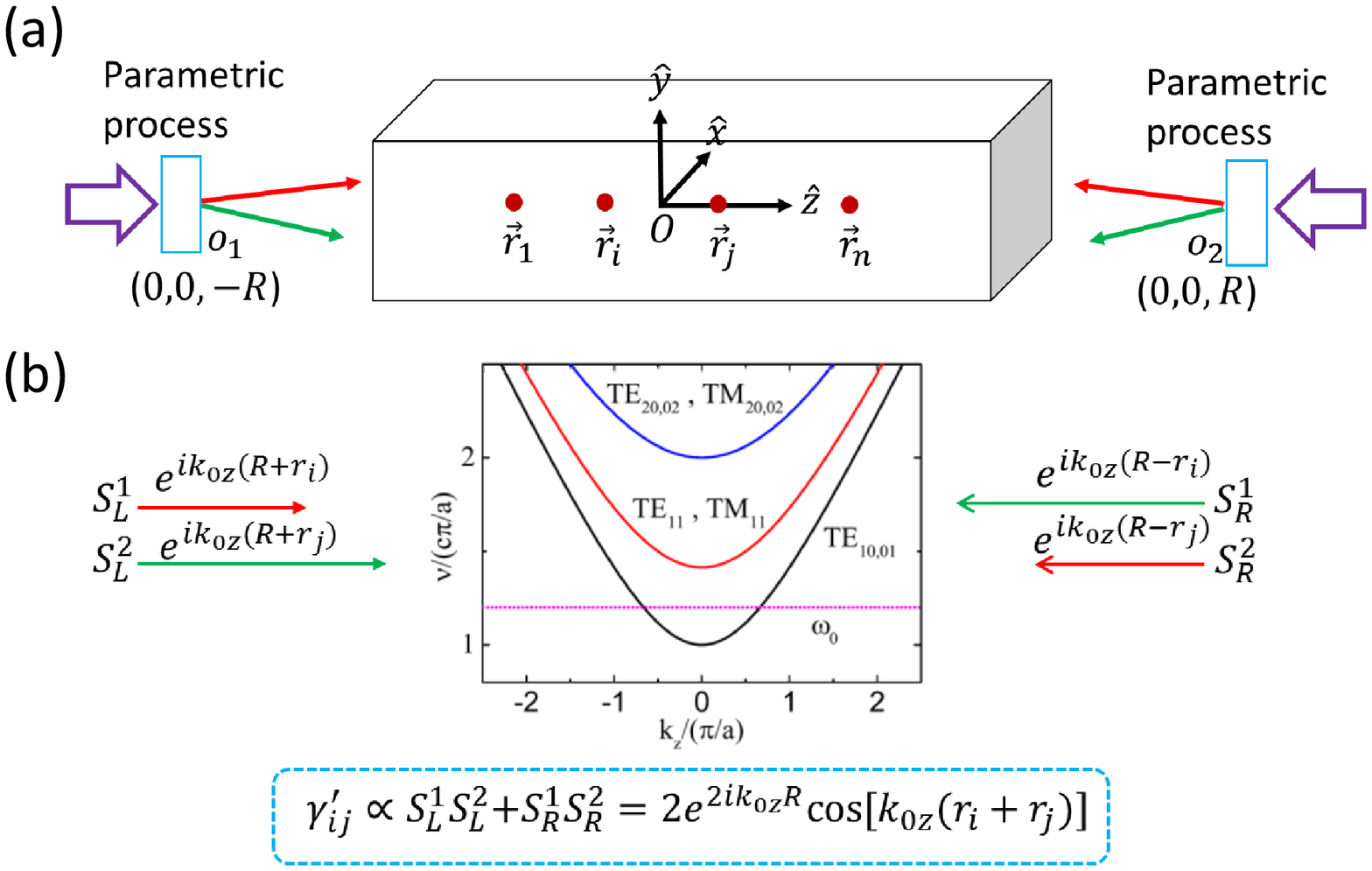}
\caption{(a) Schematic setup for waveguide-QED in a 1D squeezed vacuum where the vacuum is squeezed from both directions. (b) The dispersion relations inside the waveguide. Here the atomic transition frequency is $\frac{1.2c\pi}{a}$, which is below the cut-off frequency of $TE_{11}$ mode. Considering the fact that the atomic dipole moment is along $y$-axis and $E_{y}\ne 0$ only for $TE_{10}$, we only need to consider $TE_{10}$ mode in our calculation.}
\label{1}
\end{figure}

In practice, it is very difficult to squeeze all photon modes in 3D case. Since squeezing in 1D is experimentally achievable \cite{Murch2013, Toyli2016}, in this section we discuss the dynamics of the waveguide-QED in the squeezed vacuum. Here, we consider a perfect rectangular waveguide with negligible loss out of the waveguide as is shown in Fig.~\ref{1}(a). We assume that the cross section of the waveguide is a square with dimensions $a\times b$. The origin of the coordinate system is chosen to be at the center of the two squeezing sources with the positions of the sources to be $(0,0,\pm R)$. The emitters are located along the longitudinal centerline of the waveguide at $(0,0,r_{i})$ ($i=1,2,\cdots,N_{a}$) with the squeezed vacuum injected from both ends by the parametric process. Compared with the 3D case, the master equation in the 1D case is the same as Eq. \eqref{eq6} except that the values of $\gamma_{ij}, \gamma'_{ij}, \Lambda_{ij}$ are different.

Different from the free-space case, the square waveguide can only support certain photon modes. The allowed TE and TM modes are shown in Appendix C and their dispersion relations are shown in Fig.~\ref{1}(b). To simplify the problem, we assume that the transtion dipole moment of the emitter is along the y direction and the size of the waveguide satisfies $\lambda_{0}/2<a<\lambda_0/\sqrt{2}$ where $\lambda_{0}=2\pi c/\omega_{0}$ with $\omega_{0}$ being the transition frequency of the emitter. In this case, the emitter is mainly coupled to the $TE_{10}$ mode (Fig.~\ref{1}(b)).  The density of states of EM field in the waveguide is $D(\nu)=\frac{L}{\pi c^{2}}\frac{\nu}{\sqrt{(\frac{\nu}{c})^{2}-(\frac{\pi}{a})^{2}}}$.  The coupling strength between the emitter and the $TE_{10}$ mode is therefore given by  $g\equiv \vec{\mu}\cdot\vec{E}/\hbar=\mu\sqrt{\nu/\epsilon_{0}LS\hbar}$ \cite{Kim2013}. The single emitter decay rate due to the waveguide modes is 
\begin{equation}
\label{eq7}
\begin{split}
\gamma_{1d}=2\pi\underset{\nu}{\sum}|g(\nu)|^{2}\delta(\omega_{0}-\nu)=\frac{2\mu^{2}\omega_{0}^{2}}{\hbar\epsilon_{0}Sc^{2}k_{0z}}\equiv \eta \gamma_0,
\end{split}
\end{equation}
where $\eta=3\lambda_{0}\lambda_{0z}/(2\pi a^2)$ is the enhancement factor, $\lambda_{0z}=2\pi/k_{0z}$ is the effective longitudinal wavelength and $\gamma_0$ is the spontaneous decay rate in the free space. Around the cutoff frequency, we have $k_{0z}\rightarrow 0$ and therefore $\eta\rightarrow \infty$, i.e., the spontaneous decay rate can be greatly enhanced.

The master equation in the 1D waveguide is also given by Eq. \eqref{eq6}, but the coefficients are replaced by (see Appendix C for detail calculations):
\begin{equation}
\label{eq8}
\begin{split}
& \gamma_{ij}=\gamma_{1d}\cos(k_{0z}r_{ij}) \\
& \Lambda_{ij}=\frac{\gamma_{1d}}{2}\sin(k_{0z}r_{ij})\\
& \gamma'_{ij}=\gamma_{1d}\cos[k_{0z}(r_{i}+r_{j})]
\end{split}
\end{equation}
where $k_{0z}=\sqrt{(\frac{\omega_{0}}{c})^{2}-(\frac{c\pi}{a})^{2}}$ is the wave vector along the waveguide direction and $r_{ij}=|r_i-r_j|$ is the separation between two emitters. It is worth noting that Eq. \eqref{eq6} is valid not only for the rectangular waveguide, but also for arbitrary type of waveguide with arbitrary atomic transition frequency. The only difference for different types of waveguide and different transition frequency is the value of $\gamma_{1d}$ in Eq.~\eqref{eq7}.

Similar to the 3D case, the two-photon decay rate induced by the squeezed vacuum depends on the center of mass of the emitters. This can be explained by the interference shown in Fig.~\ref{1}(b). The emitters can  absorb two photons from the squeezing sources either from the left or the right. These two processes can interfere with each others and we have $\gamma_{ij}^{'}\propto S_{L}^{1}S_{L}^{2}+S_{R}^{1}S_{R}^{2}=2e^{2ik_{0z}R}\cos[k_{0z}(r_{i}+r_{j})]$ which is a periodic function with period $\lambda_{0z}$. Thus, when the center of mass happens to be at the antinodes (nodes) of the standing wave, the two-photon decay rate is maximized (minimized).

\subsection{One Emitter}

\begin{figure*}
\includegraphics[width=0.9\columnwidth]{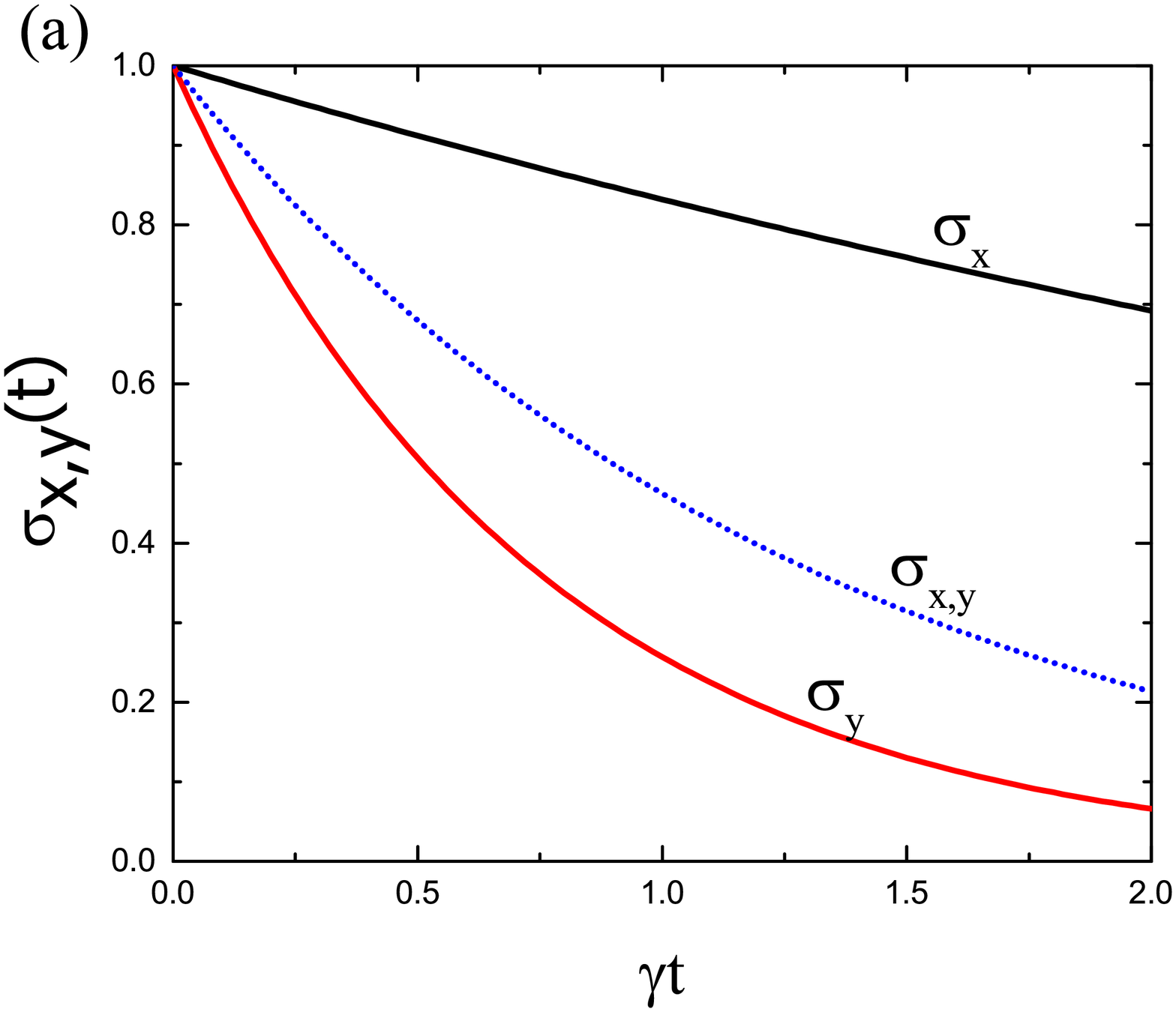}
\includegraphics[width=0.9\columnwidth,height=2.4in]{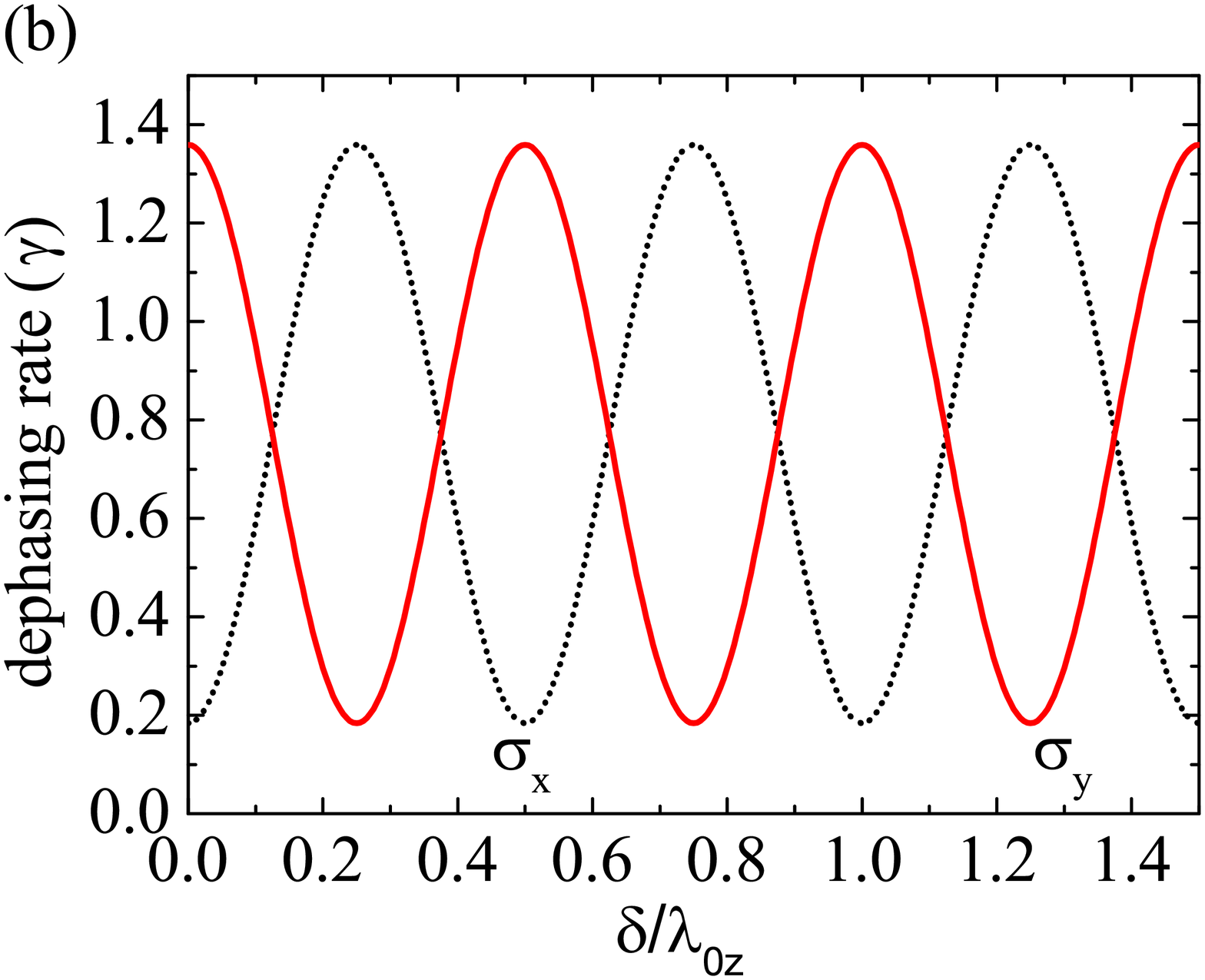}
\caption{(a) The dephasing dynamics of a single emitter in the squeezed vacuum. The black and red solid curves are the results of $\sigma_{x}$ and $\sigma_{y}$, respectively. The blue dotted line is the result when there is no squeezing (thermal reservoir). (b) The dephasing rates of $\sigma_{x}$ and $\sigma_{y}$ as a function of the emitter position. For (a)\&(b), the squeezing parameters are chosen to be $r=0.5$. 
}
\label{2}
\end{figure*}

Our theory can be used to calculate the dynamics of arbitrary number of emitters. Let us first see the one-emitter case. We still assume that the emitter is located at $(0,0,\delta)$, with the transition dipole moment along the $y$-axis. By eliminating the terms with $i\neq j$, the master equation shown in Eq.\eqref{eq6} is reduced to the single-atom case which is given by
\begin{equation}
\label{eq9}
\begin{split}
\frac{d\rho^{S}}{dt}&=\sinh(r)\cosh(r)\gamma'(e^{2ik_{0z}R}S^{+}\rho^{S}S^{+}+H.c.)\\
&-\frac{1}{2}\gamma\cosh^{2}(r)(\rho^{S}S^{+}S^{-}+S^{+}S^{-}\rho^{S}-2S^{-}\rho^{S}S^{+})\\
&-\frac{1}{2}\gamma \sinh^{2}(r)(\rho^{S}S^{-}S^{+}+S^{-}S^{+}\rho^{S}-2S^{+}\rho^{S}S^{-})\\
\end{split}
\end{equation}
with $\gamma=\gamma_{1d}$ and $\gamma'=\gamma_{1d}\cos(2k_{0}\delta)$. It is worth noting that the squeezing terms like $S^{+}\rho^{S}S^{+}$ and $S^{-}\rho^{S}S^{-}$ in Eq.~\eqref{eq9} only affect the non-diagonal terms but not the diagonal terms. Thus, for single emitter, the squeezing can only modify the dephasing rate rather than the population decay rate. We also notice that the dephasing rate due to the squeezed vacuum is dependent on the emitter position because the interference between the two squeezing sources generates a standing wave.

The dynamical equations for the expectation value of $\sigma_{+}$ and $\sigma_{-}$ are given by
\begin{equation}
\begin{split}
\frac{d}{dt}\left(\begin{array}{c}
\left\langle \sigma_{+}\right\rangle \\
\left\langle \sigma_{-}\right\rangle 
\end{array}\right)=U\left(\begin{array}{c}
\left\langle \sigma_{+}\right\rangle \\
\left\langle \sigma_{-}\right\rangle 
\end{array}\right)
\end{split}
\end{equation}
where
\begin{equation}
\begin{split}
U=\left(\begin{array}{cc}
-(N+\frac{1}{2}) & Me^{-2ik_{0z}R}\cos(2k_{0z}\delta)\\
Me^{2ik_{0z}R}\cos(2k_{0z}\delta) & -(N+\frac{1}{2})
\end{array}\right).
\end{split}
\end{equation}
The eigenvalues of $U$ are $\gamma_{dp,\pm}=N+\frac{1}{2}\pm M\cos(2k_{0z}\delta)$ which are the dephasing rate. In fact, such a position-dependent property of the dephasing rate can be associated with the variance in the quadrature phases of the squeezed field at the site of the atom. Considering the operator $X(\delta,\alpha,\beta)=\frac{1}{2\sqrt{2}}(e^{i(k_{0z}+k_{z})\delta}a_{k_{0z}+k_{z}}e^{i\alpha}+e^{i(k_{0z}-k_{z})\delta}a_{k_{0z}-k_{z}}e^{i\beta}+H.c.)$ which describes the entangled modes of the two-mode squeezing, we can find its variance $\Delta X(\delta,\alpha,\beta)=\frac{1}{2}[N+\frac{1}{2}-M\cos(2k_{0z}\delta+\alpha+\beta)]$. Therefore, we have the relation that $\gamma_{dp,+}=2\Delta X(\delta,\alpha+\beta=0)$ and $\gamma_{dp,-}=2\Delta X(\delta,\alpha+\beta=\pi)$.

We can see that when there is no squeezing, i.e., $M=0$, both $\sigma_{x}$ and $\sigma_{y}$ have the same dephasing rate $\cosh^2(r)\gamma_{1d}/2$ (blue dotted line in Fig.~\ref{2}(a)). However, if there is squeezing, i.e., $M\neq 0$, $\sigma_{x}$ and $\sigma_{y}$ have different dephasing rates with one being enhanced and the other one being suppressed (solid lines in Fig.~\ref{2}(a)).
The dephasing rate can be tuned by changing the position of the emitter. In Fig.~\ref{2}(b), it is shown that the dephasing rates of $\sigma_{x}$ and $\sigma_{y}$ vary periodically as the emitter position changes. At some regions, $\sigma_{x}$ decays faster than $\sigma_{y}$, while at other regions, $\sigma_{x}$ decays slower than  $\sigma_{y}$. This result challenges the traditional conclusion where dephasing rate is a position-independent constant\cite{Zubairy1988, Scully1997}.

The power spectrum of the resonance fluorescence can also be calculated and the result is similar to Ref. \cite{Carmichael1987} with the simple replacements of $M$ by $M\gamma'$ and the phase of $M$ by $e^{2ik_{0z}R}$.

\subsection{Two Emitters}

\begin{figure*}
\includegraphics[width=0.9\columnwidth]{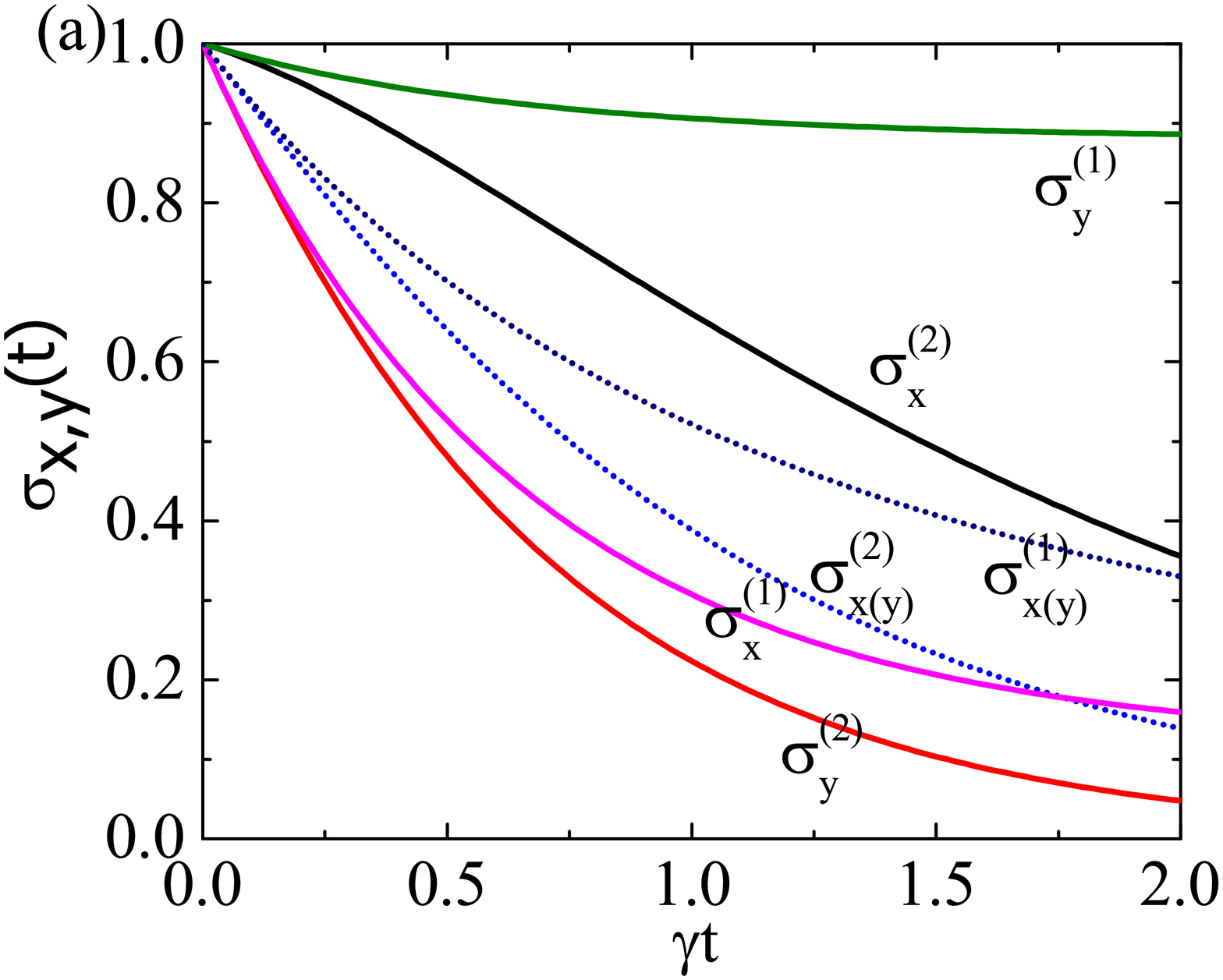}
\includegraphics[width=0.9\columnwidth]{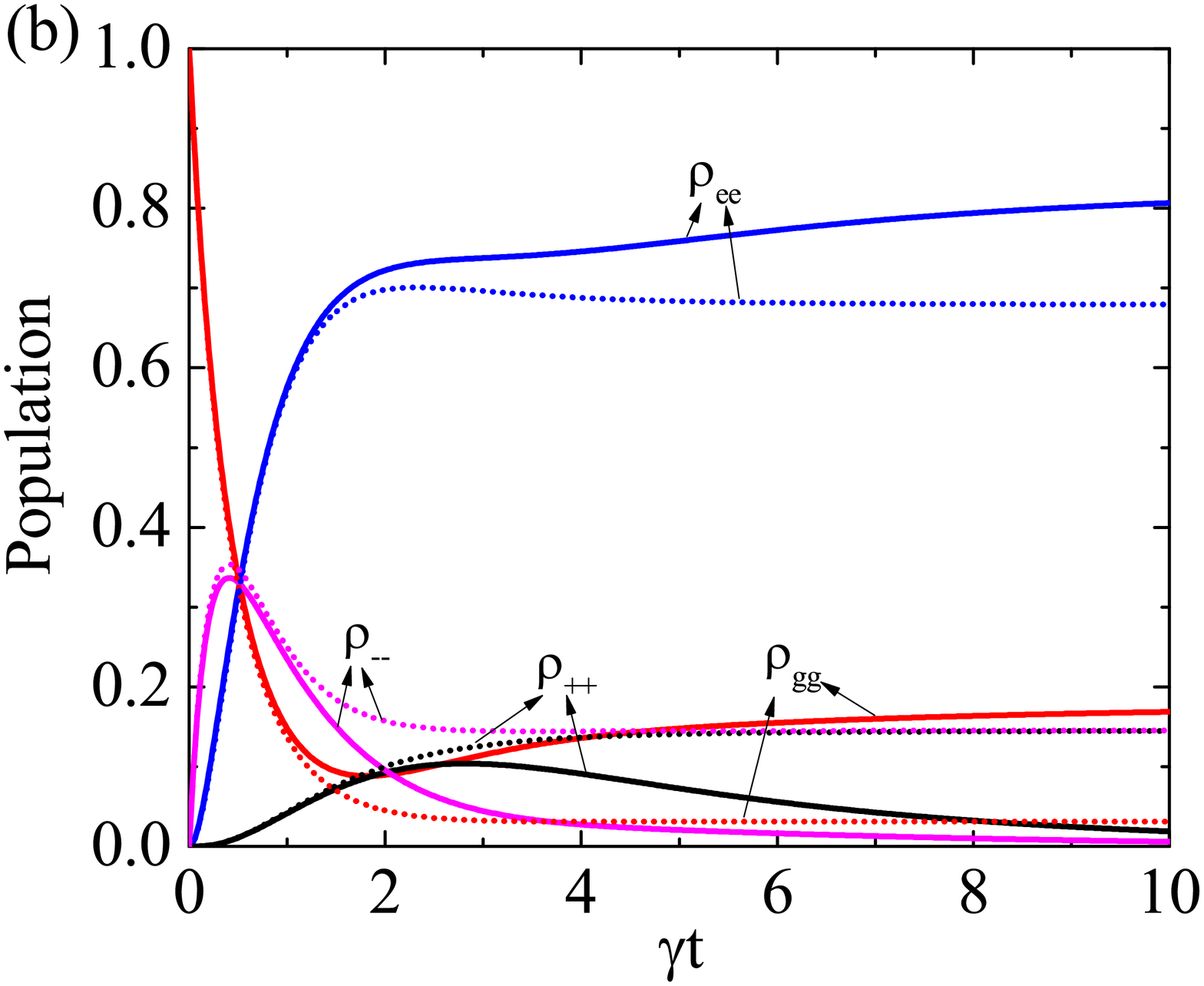}
\includegraphics[width=0.9\columnwidth]{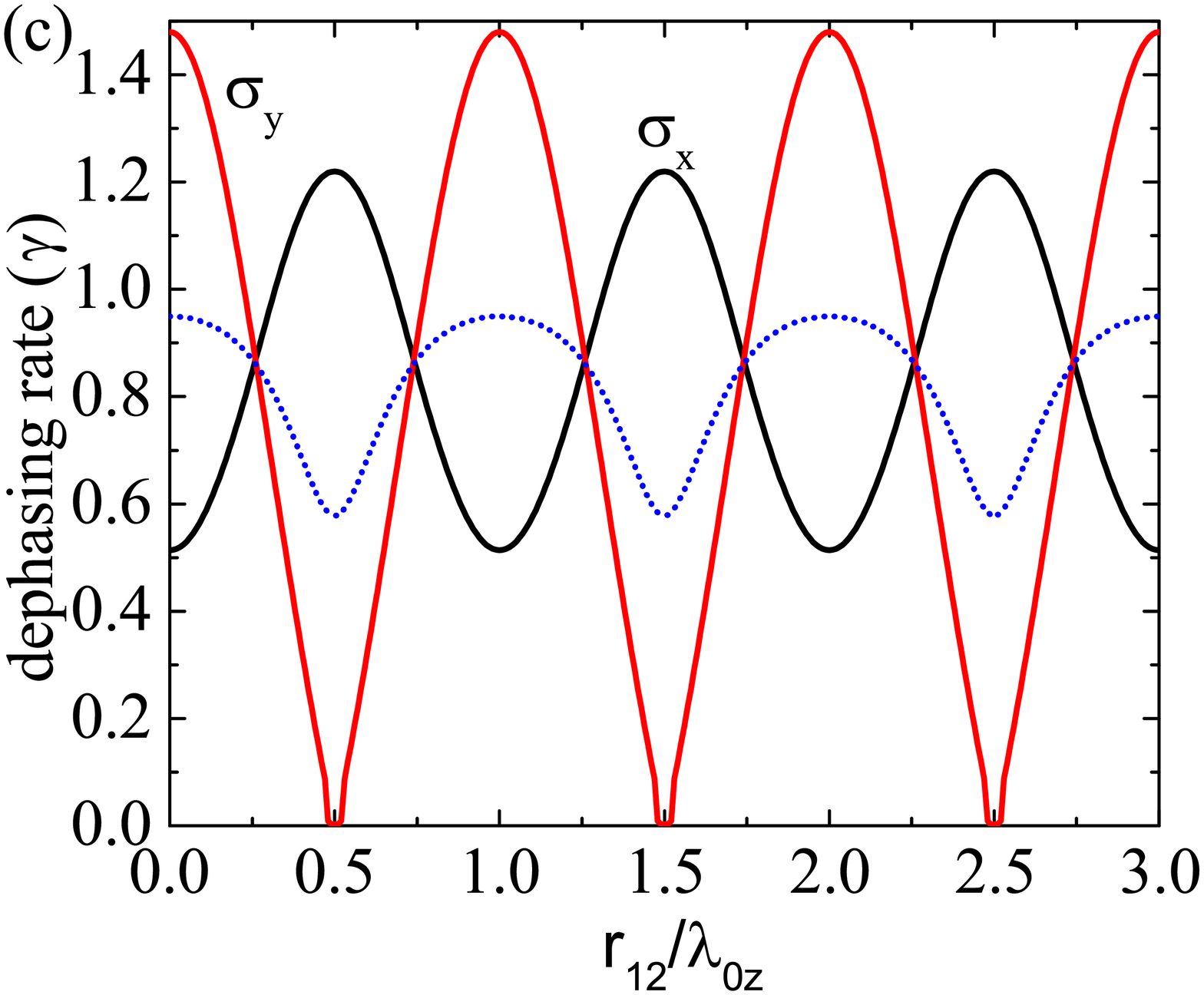}
\includegraphics[width=0.9\columnwidth,height=2.4in]{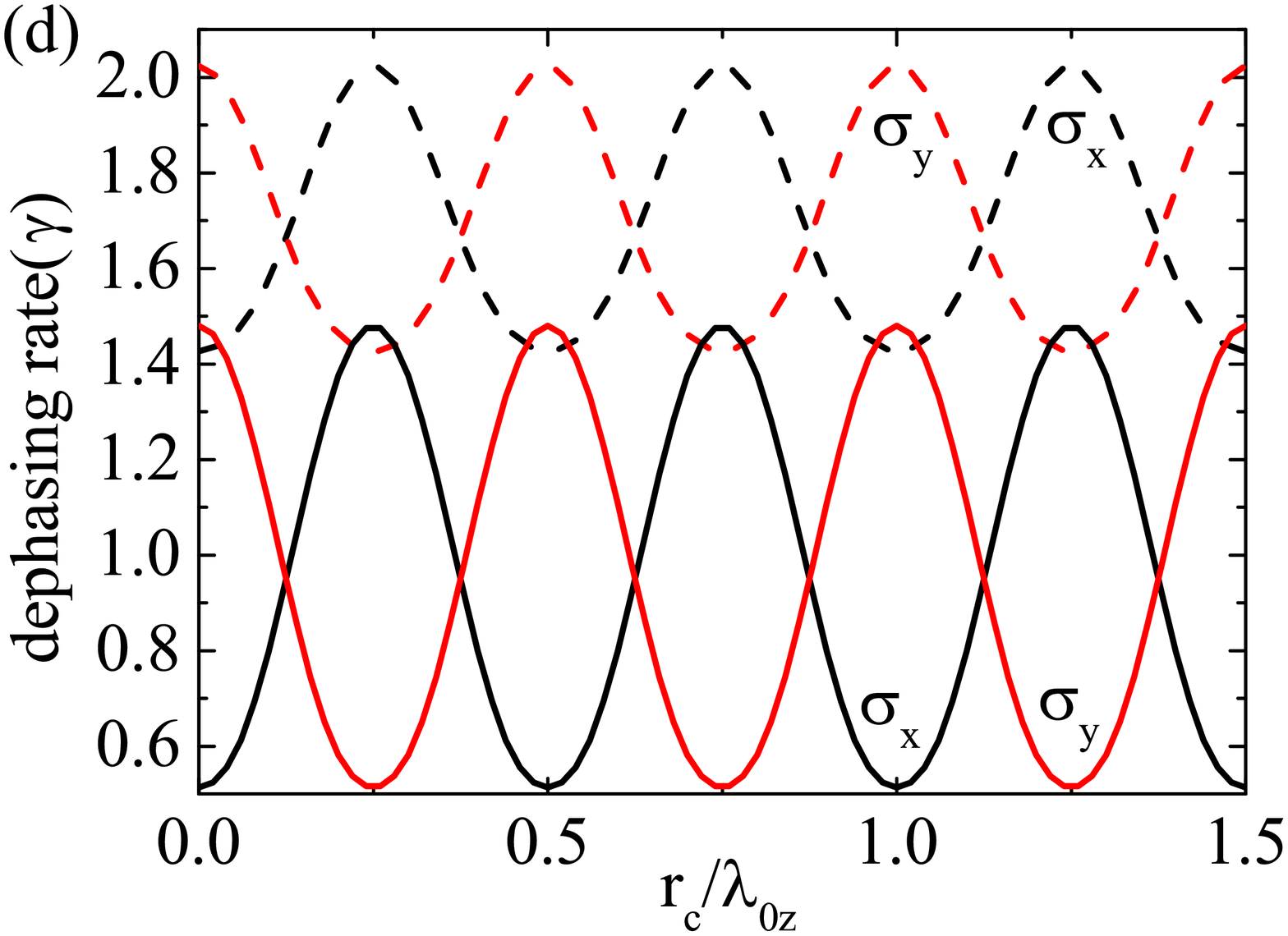}

\caption{Two-emitter case: Transverse polarization decay of the first emitter as a function of time. (a) $r_{12}=0.5\lambda_{0z}$ for superscript (1) and $r_{12}=1.0\lambda_{0z}$ for superscript (2), $r=0.5$ and $r_{c}=0$; (b)  population decay as a function of time when $r_{12}=0.5\lambda_{0z}$, $r=0.5$ and $r_{c}=0$. Solid lines are the results in squeezed vacuum and the dotted lines are the results in the thermal reservoir with $N=\sinh^2(r)$. Here the dynamics of $\rho_{++}$ and $\rho_{--}$ are highly identical. (c) Dephasing rate as a function of atom separation with the center of mass fixed at $r_{c}=0$. (d) Dephasing rate as a function of center of mass position with atom separation fixed at $r_{12}=\lambda_{0z}$, where the two-atom case is plotted in solid lines and the five-atom case is plotted in dashed lines.}
\label{3}
\end{figure*}

Next, we consider the two-emitter case where dipole-dipole interaction can occur and two-photon process is allowed.  In Fig.~\ref{3}(a), we show the dynamics of the transverse polarization $\sigma_{x}$ and  $\sigma_{y}$. Here, we compare two different emitter separations $r_{12}=0.5\lambda_{0z}$ and $r_{12}=1.0\lambda_{0z}$. In both cases, the $x$ and $y$ polarizations have the same decay dynamics in the thermal reservoir. However, in the squeezed vacuum, the two orthogonal polarizations have different decay rates with one being enhanced and the other being suppressed. When $r_{12}=0.5\lambda_{0z}$, $\sigma_{x}$ decays faster than that in the thermal reservoir, but  $\sigma_{y}$ decays much slower than that in the thermal reservoir. While opposite result occurs when $r_{12}=1.0\lambda_{0z}$. This is similar to the one-emitter case. 

Different from the one-emitter case, as is shown in Fig.~\ref{3}(b), the squeezed vacuum can affect the population decay of the two-emitter system. This is because two-photon process is allowed in the two-emitter system. Without the squeezed vacuum, the system is finally in the thermal equilibrium state (dotted lines). However, the squeezed vacuum can deplete the populations on $|++\rangle$ and $|--\rangle$ with $|\pm\rangle=\frac{1}{\sqrt{2}}(|e_{1}\rangle|g_{2}\rangle \pm |g_{1}\rangle|e_{2}\rangle)$. In fact, the atomic pair evolves into an entanglement state in this case and we will discuss it later.

We also study the dephasing rate as a function of emitter separation and position of the center of mass which are shown in Fig.~\ref{3}(c) and (d) respectively. Here the dephasing rate is defined to be the inverse of time for $\sigma_x(\sigma_y)$ to damp to $1/e$ of its initial value. Similar to the one-emitter case, the dephasing rate is a periodic function of both $r_{12}$ and $r_{c}$. However, due to the dipole-dipole interaction, the dephasing rate is no longer a constant even in the thermal reservoir (dotted line in Fig.~\ref{3}(c)) so that the value ranges of $\sigma_x$ and $\sigma_y$ are no longer the same in the squeezed vacuum(solid lines in Fig.~\ref{3}(c)). It is noted that when $r_{12}=0.5n\lambda_{0z}$ ($n$ is any integer) $\sigma_{y}$ does not decay to $1/e$ of its initial value due to the subradiance effect. When we fix the atom separation and change the center of mass(Fig.~\ref{3}(d)), the dephasing rate changes periodically and harmonically like one-emitter case.  Therefore, the dephasing rate is tunable by changing the atom separation or position of center of mass. Usually, the positions of the atoms are not easy to be tuned. However, we can easily tune the position of the squeezing sources to effectively change the center of mass of the atoms. Figure~\ref{3}(d) also shows the result when there are five emitters (dashed lines). The dephasing rate is significantly increased when $N_a$ increases due to the collective effect.

\subsection{Quantum Entanglement}

Quantum entanglement is an important resource of the quantum information and quantum metrology \cite{Horodecki2009, Giovannetti2011}. Preparation of the maximum entangled state is still a central topic of interest. It has been shown that stationary quantum entanglement can be dissipatively prepared by engineering the bath enviroment \cite{Kraus2008, Diehl2008, Lin2013, Ma2015}. By squeezing the enviroment, quantum entanglement between emitters can be also created \cite{Kraus2004, Tanas2004, Li2006}. However, it is shown in Ref. \cite{Tanas2004} that stationary maximum entanglement can not be reached by the squeezed vacuum for identical emitters. Here, we show that identical emitters coupled to the 1D waveguide can also be driven to a stationary maximum entangled NOON state by the squeezed vacuum as long as the center of mass is put at the proper position.

\begin{figure*}
\includegraphics[width=0.8\columnwidth]{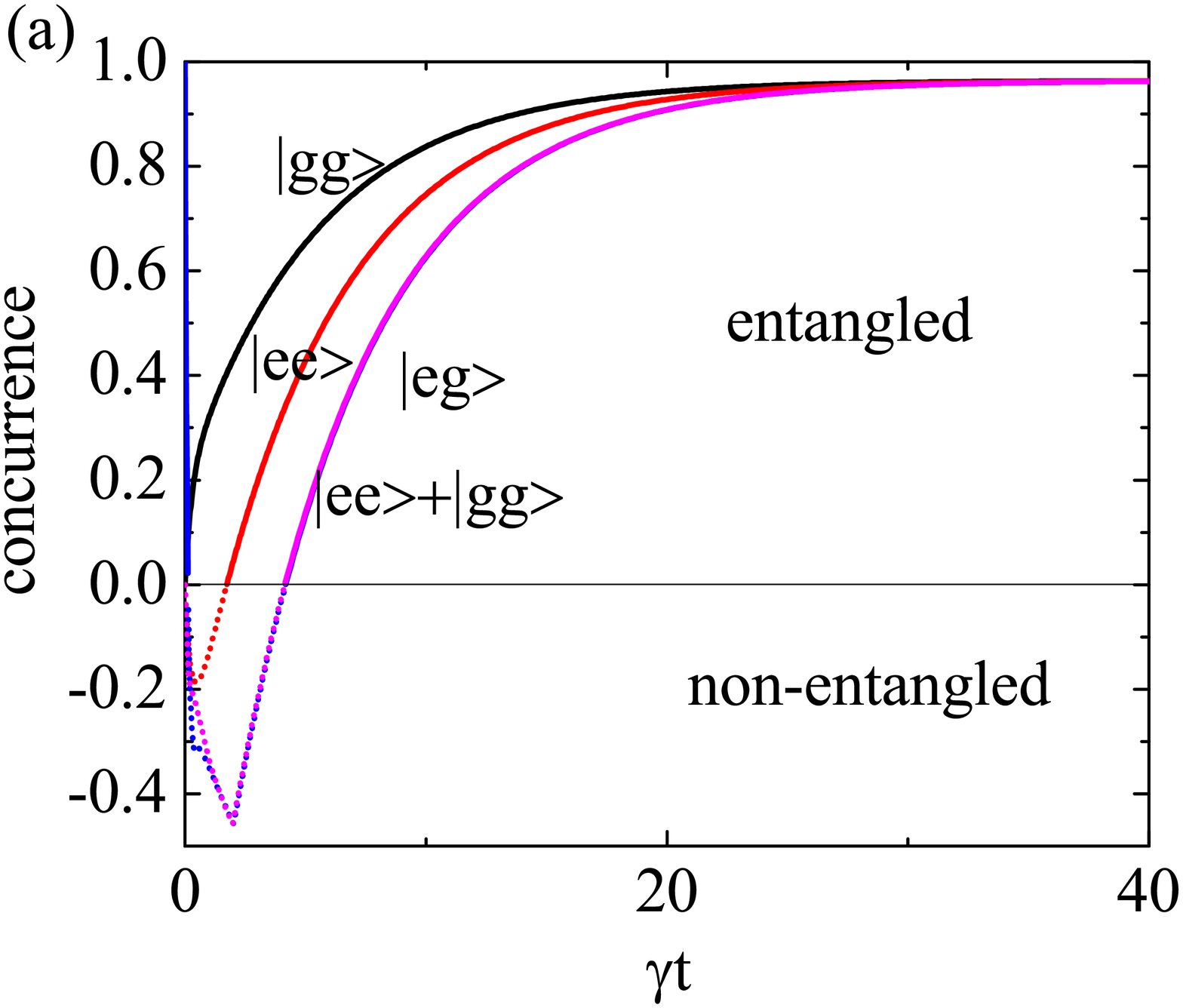}
\includegraphics[width=0.8\columnwidth]{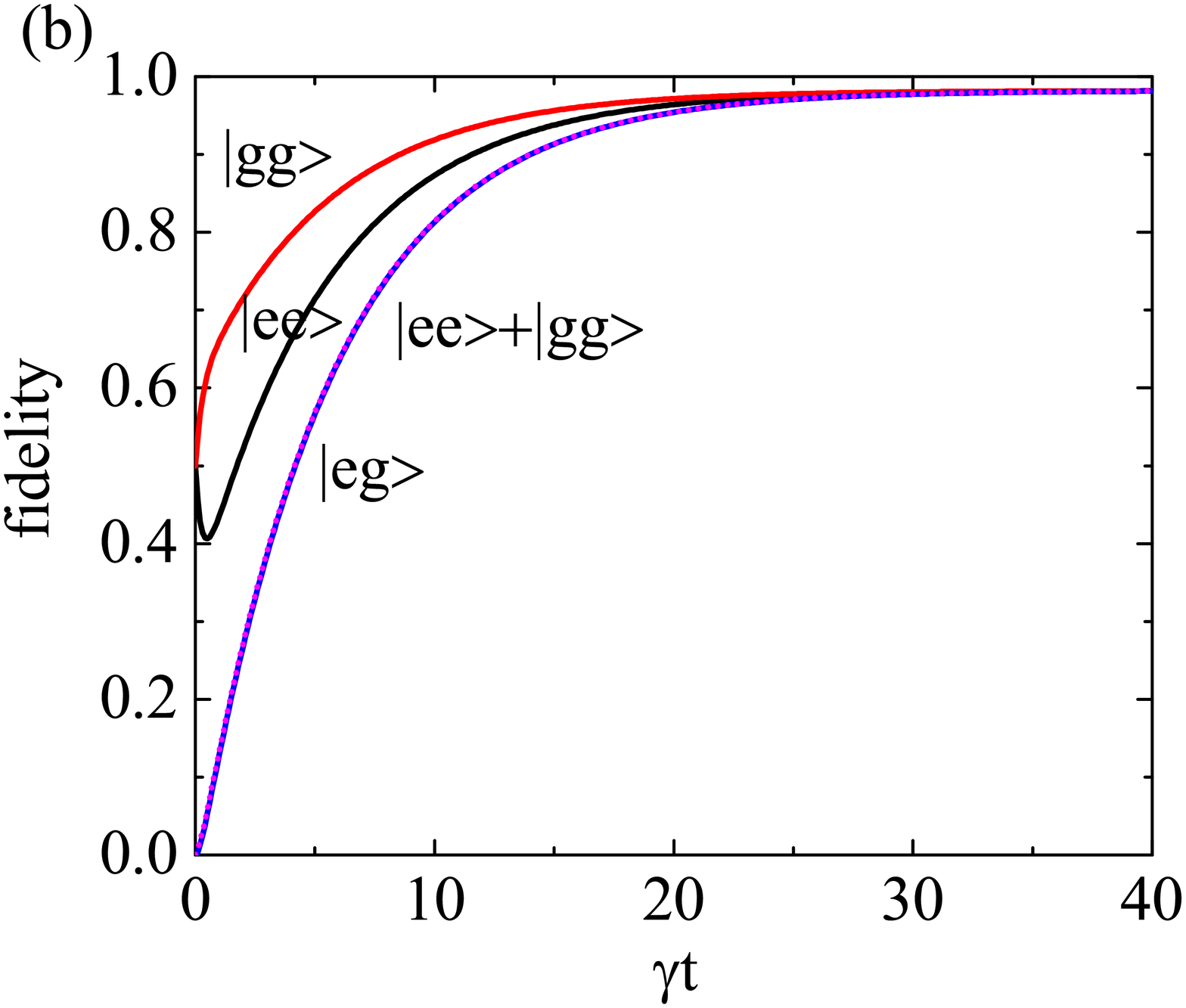}
\caption{(a) Concurrence evolution of different initial states in squeezed vacuum, where $r=1$, $r_c=0$, and $r_{12}=0.25\lambda_{0z}$. (b) Fidelity evolution of different initial states in the same environment.}
\label{4}
\end{figure*}

The quantum entanglement can be measured by the concurrence which is defined as \cite{Hill1997}: $\mathscr{C}\equiv max\{0,\lambda_{1}-\lambda_{2}-\lambda_{3}-\lambda_{4}\}$ in which $\lambda_{1},\lambda_{2},\lambda_{3},\lambda_{4}$ are eigenvalues, in decreasing order, of the Hermitian matrix $R=\sqrt{\sqrt{\rho}\widetilde{\rho}\sqrt{\rho}}$ with $\widetilde{\rho}=(\sigma_{y}\bigotimes\sigma_{y})\rho^{\ast}(\sigma_{y}\bigotimes\sigma_{y})$. For a pure two-qubit state $|\Psi\rangle=\alpha |ee\rangle +\beta |eg\rangle+\gamma |ge\rangle +|gg\rangle$ with $|\alpha|^{2}+|\beta|^{2}+|\gamma|^{2}+|\delta|^{2}=1$, the concurrence is given by $\mathscr{C}=max\{0,2|\alpha\delta-\beta\gamma|\}$. The concurrence as a function of time for different initial states is shown in Fig.~\ref{4}(a) where $r=1, r_c=0,$ and $r_{12}=0.25\lambda_{0z}$. Different curves correspond to different initial states. We can see that no matter what the initial state is, the two-emitter state will be driven to a very high entangled state. To see what the stationary state is, we also show the fidelity of the emitter state with respect to the maximum entangled state $\frac{1}{\sqrt{2}}(|gg\rangle -|ee\rangle)$ which is shown in Fig.~\ref{4}(b). We can see that the stationary state is very close to it. Therefore, under these parameters the two emitters can be driven to the maximum entangled state which may find important applications in quantum information and quantum computation.

To find the stationary state analytically, we rewrite the master equation in Eq.~\eqref{eq6} as
\begin{eqnarray}
\dot{\rho}_{gg}&=&-2N\gamma\rho_{gg}+(N+1)\gamma_{+}\rho_{++}+(N+1)\gamma_{-}\rho_{--}\nonumber \\ & & +M\gamma^{'}_{12}\rho_{u},  \\
\dot{\rho}_{ee}&=&-2(N+1)\gamma\rho_{ee}+N\gamma_{+}\rho_{++}+N\gamma_{-}\rho_{--}\nonumber \\ & &+M\gamma^{'}_{12}\rho_{u},  \\
\dot{\rho}_{++}&=&-(2N+1)\gamma_{+}\rho_{++}+(N+1)\gamma_{+}\rho_{ee}+N\gamma_{+}\rho_{gg}\nonumber \\ & &-M\gamma^{'}_{+}\rho_{u},  \\
\dot{\rho}_{--}&=&-(2N+1)\gamma_{-}\rho_{--}+(N+1)\gamma_{-}\rho_{ee}+N\gamma_{-}\rho_{gg}\nonumber \\ & &-M\gamma^{'}_{-}\rho_{u}. \\
\dot{\rho}_{u}&=&-(2N+1)\gamma_{11}\rho_{u}-2M\gamma^{'}_{+}\rho_{++}-2M\gamma^{'}_{-}\rho_{--} \nonumber \\ & & +2M\gamma^{'}_{12}(\rho_{ee}+\rho_{gg}).
\end{eqnarray}
where $\rho_{ee}=\langle ee|\rho |ee\rangle$, $\rho_{gg}=\langle gg|\rho |gg\rangle$, $\rho_{\pm\pm}=\langle \pm|\rho |\pm\rangle$ with $|\pm\rangle=\frac{1}{\sqrt{2}}(|e_{1}\rangle|g_{2}\rangle \pm |g_{1}\rangle|e_{2}\rangle)$, $\rho_{u}=e^{-2ik_{0z}R}\langle ee|\rho |gg\rangle+e^{2ik_{0z}R}\langle gg|\rho |ee\rangle$, and $\gamma=\gamma_{1d}$, $\gamma_{\pm}=\gamma_{1d}(1\pm\cos(k_{0z}r_{12}))$, $\gamma_{12}^{'}=\gamma_{1d}\cos(2k_{0z}r_{c})$, $\gamma^{'}_{\pm}=\gamma_{1d}\{\cos[2k_{0z}r_{c}]\pm\frac{1}{2}[\cos(2k_{0z}r_{1})+\cos(2k_{0z}r_{2})]\}$ with $r_{c}=\frac{(r_{1}+r_{2})}{2}$. Then the steady state solutions are given by
\begin{equation}
\label{eq11}
\begin{split}
&\rho_{ee}=\frac{N[-1-N-2N^{2}+(-1+N+2N^{2})\cos(4k_{0z}r_{c})]}{2(1+2N)[-1-2N-2N^{2}+2N(1+N)\cos(4k_{0z}r_{c})]}\\
&\rho_{++}=-\frac{N(1+N)sin^{2}(2k_{0z}r_{c})}{-1-2N-2N^{2}+2N(1+N)\cos(4k_{0z}r_{c})}\\
&\rho_{--}=-\frac{N(1+N)sin^{2}(2k_{0z}r_{c})}{-1-2N-2N^{2}+2N(1+N)\cos(4k_{0z}r_{c})}\\
&\rho_{u}=\frac{-2\sqrt{N(1+N)}\cos(2k_{0z}r_{c})}{(1+2N)[-1-2N-2N^{2}+2N(1+N)\cos(4k_{0z}r_{c})]}\\
\end{split}
\end{equation}
where we have used the relation $M^2=N(N+1)$. Obviously, the population given by Eq.~\eqref{eq11} differs from that given by thermal reservoir: $\rho_{ee(gg)}=\rho^{th}_{ee(gg)}+\Delta\rho, \rho_{++(--)}=\rho^{th}_{++(--)}-\Delta\rho$ with $\Delta\rho=\frac{N(N+1)\cos^2(2k_{0z}r_c)}{(1+2N)^2(1+2N+2N^2-2N(1+N)\cos(4k_{0z}r_c))}$ and $\rho^{th}_{ee}=\frac{N^2}{(1+2N)^2}$, $\rho^{th}_{++}=\rho^{th}_{--}=\frac{N(N+1)}{(1+2N)^2}$, $\rho^{th}_{gg}=\frac{(1+N)^2}{(1+2N)^2}$ which obey the Boltzmann distribution. It is interesting that the steady state depends only on the center of mass but not on the separation between the two emitters. Meanwhile, it is worth noting that the dark state cannot always be reached since the ergodicity cannot be guaranteed under every condition. For example, when $\cos(k_{0z}r_{12})= 1$, $|+\rangle$ becomes a dark state, while it is $|-\rangle$ when $\cos(k_{0z}r_{12})=-1$. 

Eq.~\eqref{eq11} shows that as $r_c$ gets closer to $\frac{n}{4}\lambda_{0z}$, the magnitude of $\gamma'_{\pm}$ gets closer to $\pm1$ which leads to smaller population on $|+\rangle$ and $|-\rangle$ as well as bigger concurrence. When the position of the center mass $r_{c}=\frac{n}{4}\lambda_{0z}$, the steady states are given by
\begin{equation}
\label{eq12}
\begin{split}
&\rho_{gg}=\frac{N+1}{(1+2N)},\\
&\rho_{ee}=\frac{N}{(1+2N)},\\
&\rho_{++}=\rho_{--}=0,\\
&\rho_{u}=(-1)^{n+1}\frac{2\sqrt{N(1+N)}}{(1+2N)}.\\
\end{split}
\end{equation}
which corresponds to the state $|\Psi_{s}\rangle=\frac{1}{\sqrt{2N+1}}(\sqrt{N+1}|gg\rangle +(-1)^{n+1} \sqrt{N}|ee\rangle)$.  The concurrence of this state is given by $\mathscr{C}=|\rho_u|-(\rho_{++}+\rho_{--})=\frac{2\sqrt{N(N+1)}}{(2N+1)}$, which monotonically increases with the average photon number $N$. When $N\rightarrow\infty$, $\mathscr{C}\rightarrow 1$ which is a maximum-entangled state $\frac{1}{\sqrt{2}}(|gg\rangle -|ee\rangle)$ ($\frac{1}{\sqrt{2}}(|gg\rangle +|ee\rangle)$) with even(odd) $n$. 

\begin{figure}
\includegraphics[width=0.8\columnwidth]{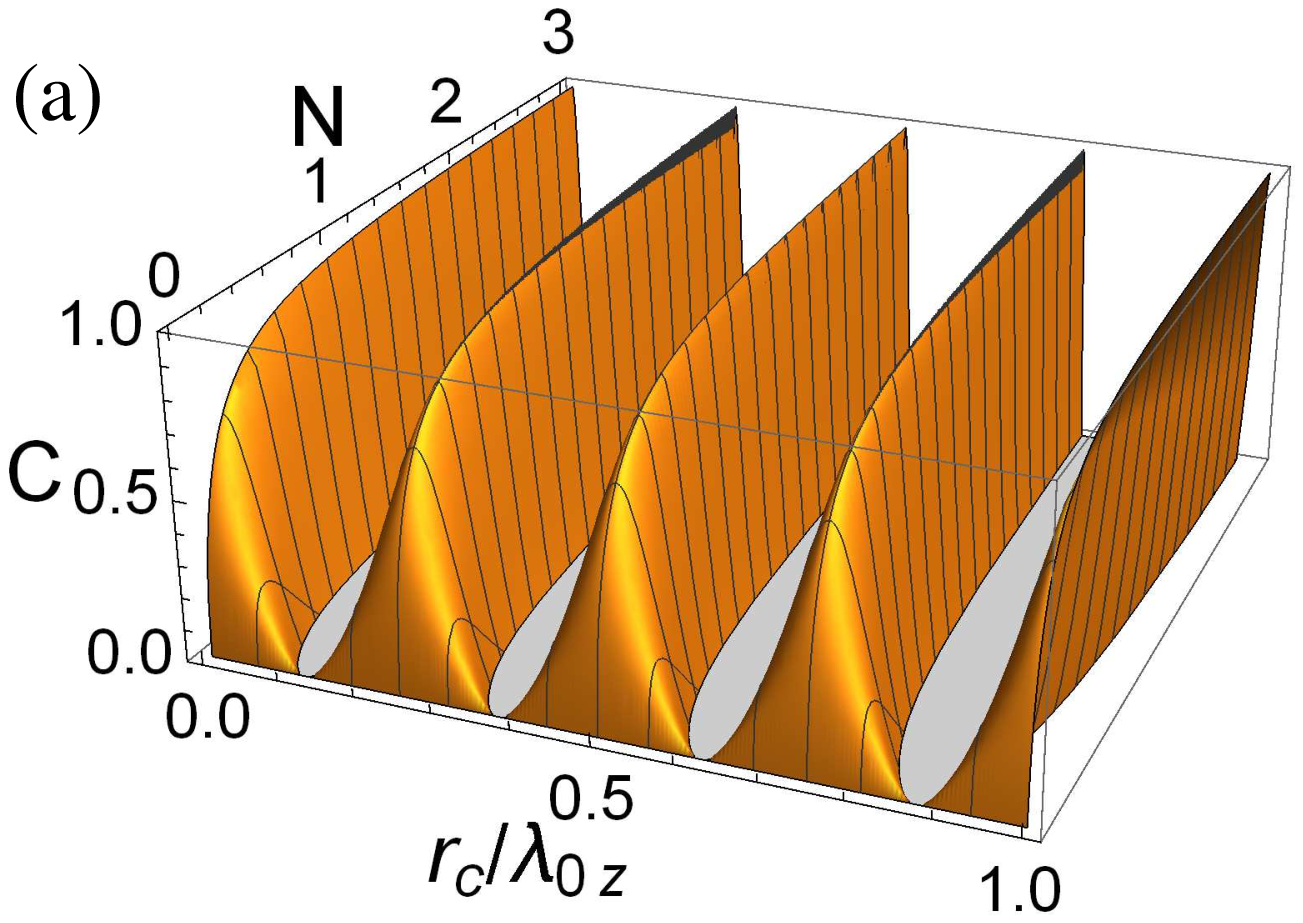}
\includegraphics[width=0.8\columnwidth]{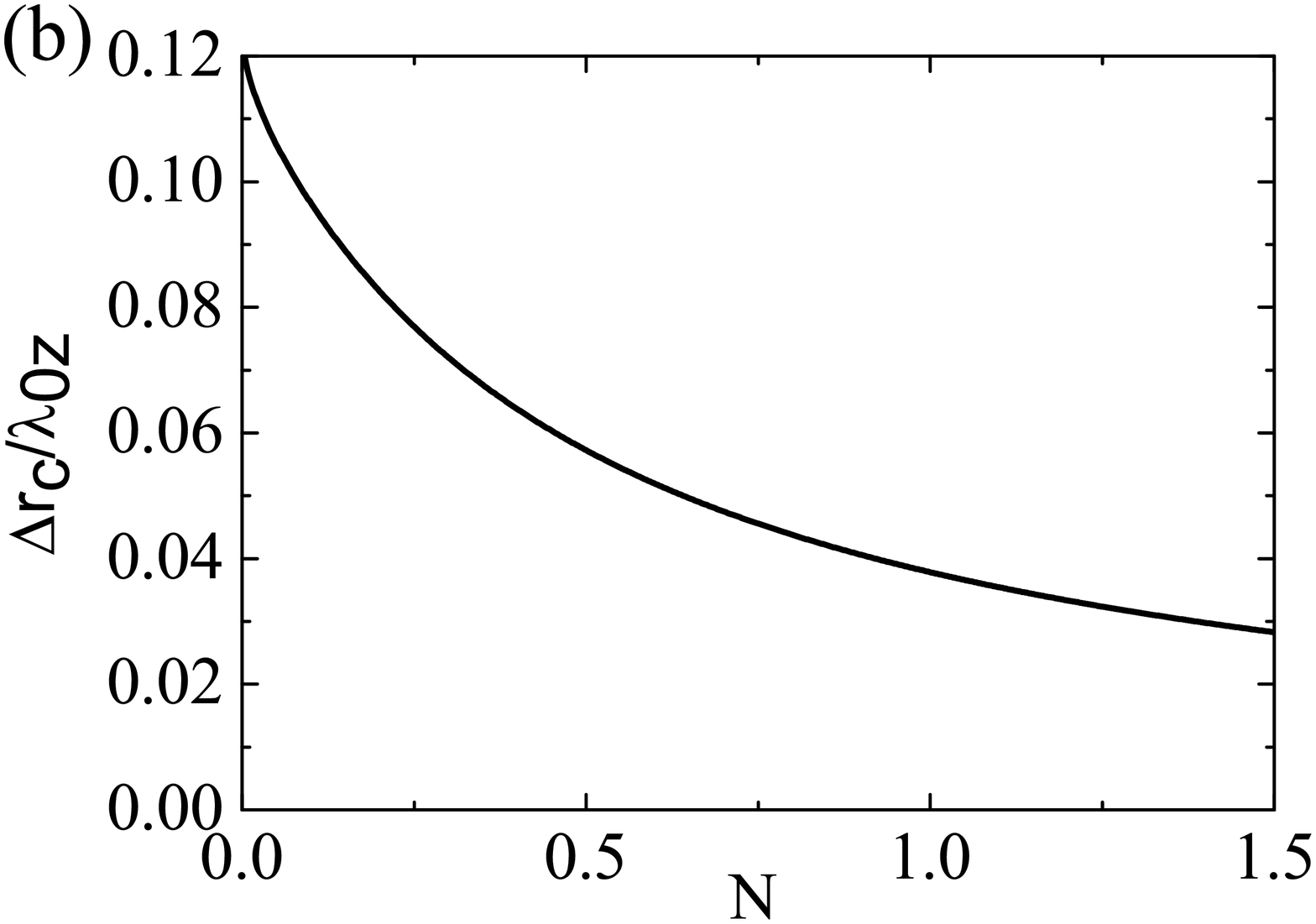}
\caption{(a) Concurrence of the steady state as a function of average photon number $N=\sinh(r)^2$ and the position of the center mass $r_c=\frac{r_1+r_2}{2}$.(b) The impact of $r_c$'s fluctuations on concurrence for different average photon number $N$. $\Delta r_c$ is the distance from $\frac{n}{4}\lambda_{0z}$ to the position where the entanglement vanishes.}
\label{5}
\end{figure}

Fig.~\ref{5}(a) shows the dependence of the stationary quantum entanglement on the photon number and the center-of-mass position. It is clearly  seen that when $r_{c}$ is close to $\frac{n}{4}\lambda_{0z}$ the system can be prepared in a high entangled state, while the entanglement can never be formed when $r_c=\frac{2n+1}{8}\lambda_{0z}$ because the dipole-dipole interaction $\gamma'_{12}$ vanishes. In experiments, the center of mass position of emitters may be hard to control, but it can be effectively controllable by setting the positions squeezing sources. Thus, as long as the pump beam in SPDC is strong enough to guarantee the average photon number of the squeezed vacuum, the emitters can definitely evolve into a NOON state. 
While the dephasing rate is not very sensitive to the fluctuations of the emitter positions, the stationary quantum entanglement significantly depends on their center of mass. Only when the center of mass position is around $n\lambda/4$, the quantum entanglement is nonzero. In Fig.~\ref{5}(b), we show half the range of center of mass where the quantum entanglement is non-zero. The larger the squeezing is, the more sensitive the quantum entanglement is to the fluctuation of center-of-mass. For example, when $N=1$, a deviation of about $0.04\lambda$ from $n\lambda/4$ will make the entanglement vanish.

\subsection{Resonance Fluorescence}

\begin{figure*}
\includegraphics[width=2\columnwidth]{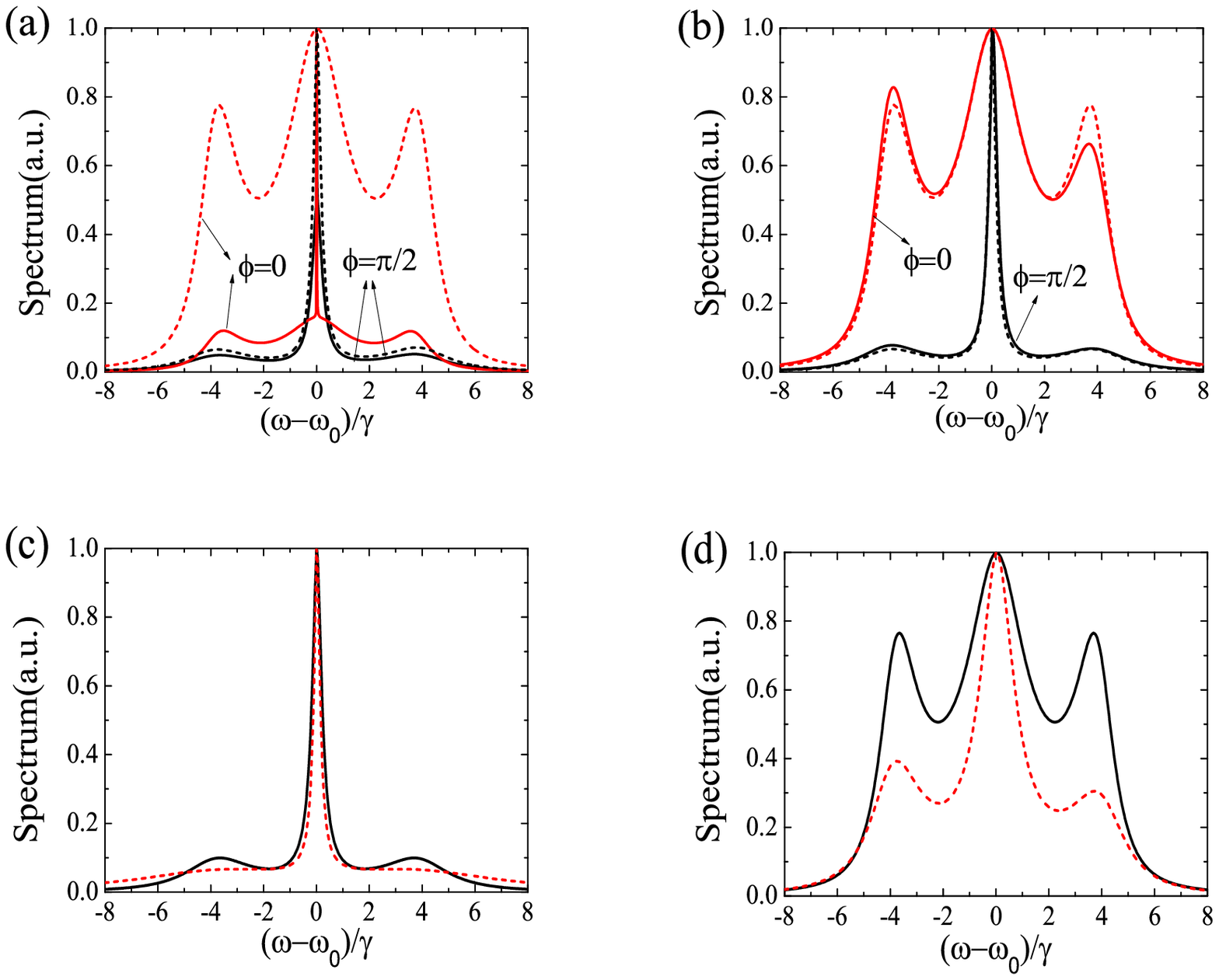}
\caption{Resonance fluorescence spectrum of the two-emitter system inside a 1D waveguide. For better comparison, the spectra are normalized to the intensity at $\omega=\omega_0$ with the coherent elastic scattering singularity removed. Coherent driving Rabi frequency is $\Omega_R=4\gamma$. In (a) and (b), the solid curves are the spectra for the coupled emitters, while the dashed curves are the spectra without emitter-emitter coupling. Parameters: (a) $r_1=0, r_2=0.01\lambda_{0z}$, squeezing parameter $r=0.5$. (b) $r_1=0, r_2=0.25\lambda_{0z}$, $r=0.5$. (c) $r_1=0, r_2=\lambda_{0z}, \phi=\pi/2$, $r=0.5$ for black line, $r=1$ for red line. (d) $r_1=-0.125\lambda_{0z}, r_2=0.125\lambda_{0z}$ for the red line, $r_1=-0.25\lambda_{0z}, r_2=0.25\lambda_{0z}$ for the black line. $\phi=0, r=0.5$.}
\label{6}
\end{figure*}

In this subsection, we study how the squeezing can affect the resonance fluorescence of the waveguide-QED system. In the following we study how the collective interaction, squeezing phase, squeezing degree, emitter separation, and the center of mass affect the resonance fluorescence of this system. 

The power spectrum of the resonance fluorescence is given by \cite{Scully1997, Ficek1990a, Liao2012}
\begin{equation}
S(\omega)\propto Re\int_{0}^{\infty}d\tau Tr[\sigma^{-}(\tau)\sigma^{+}(0)]e^{i\omega\tau}.
\end{equation}
where we assume that the detector is perpendicular to the waveguide and $\sigma^{\pm}=\sigma^{\pm}_{1}+\sigma^{\pm}_{2}$ for the two-emitter example. The two-time correlation function in the integration can be calculated by the quantum regression theorem. Usually, the analytical result of Eq. (22) is difficult to get. However, we can resort to the numerical method to calculate the resonance fluorescence \cite{Molmer1993}.

To observe the resonance fluorescence, we need to apply an external coherent driving field. The master equation is given by
\begin{equation}
\label{eq13}
\begin{split}
\frac{d\rho}{dt}=-i[V,\rho]+\mathcal{L\rho}
\end{split}
\end{equation}
where $\mathcal{L\rho}$ is the right hand side of Eq.\eqref{eq6} and $V=\frac{\Omega_{R}}{2}e^{-i\alpha}(e^{-ik_{0z}r_{1}}\sigma_{1}^{-}+e^{-ik_{0z}r_{2}}\sigma_{2}^{-})+H.c.$ is the interaction between the driving field and the emitters with Rabi frequency $\Omega_{R}=\frac{\vec{d}\cdot\vec{E}}{\hbar}$. From Eq.~\eqref{eq13} we can evolve and obtain the steady state of the system $\rho_{ss}$. Next we use $(\sigma_{1}^{-}+\sigma_{2}^{-})\rho_{ss}$ as the initial condition to solve a density matrix $c(t)$ which obeys the same equation of motion as $\rho$ in Eq.~\eqref{eq13}.  The resonance fluorescence spectrum is then given by \cite{Molmer1993}
\begin{equation}
S(\omega)\propto Re\int_{0}^{\infty}d\tau Tr[c(\tau)(\sigma_{1}^{+}+\sigma_{2}^{+})]e^{i\omega\tau}.
\end{equation}

In Fig. 6(a) and 6(b) we compare the resonance fluorescence spectrum with and without the dipole-dipole interaction for different squeezing phases and emitter separations. When $r_{12}=0.01\lambda_{0z}$ and $\phi=0$， we can see that the spectrum is very different with and without dipole-dipole interaction. Without dipole-dipole interaction, the spectrum is very similar to the typical Mollow triplet (red dashed line). However, with dipole-dipole interaction, there is a very narrow peak around the center frequency (red solid line). This is due to the subradiant state induced by the dipole-dipole interaction. On the contrary, when $\phi=\pi/2$ the spectrum with and without the dipole-dipole interaction is very similar (black solid and dashed lines).  From Fig. 6(b) we see that with dipole-dipole interaction, the spectrum can be asymmetric, i.e., the positive and negative sidebands are different.

In Fig. 6(c) we compare the spectrum with different squeezing degrees. We can see that greater squeezing parameter leads to the power spectrum in weak-driving-field limit(sidebands disappear). FIG. 6(d) shows that different emitter separation has different spectrum. This is not only due to atomic interaction which is described by $\gamma_{12},\gamma'_{12},\Lambda_{12}$, but also due to their positions which determine the values of $\gamma'_{ii}$, i.e., the effective phase and magnitude of $M$. Comparing the red solid curve in Fig. 6(b) and the red dashed curve in Fig. 6(d) we can see that different center-of-mass position can also have different resonance fluorescence.

\section{Summary}
We modify the usual squeezed vacuum mode function to include the position information of the squeezing source and derive a master equation of the atom dynamics based on the Weisskopf-Wigner approximation. In our formalism, the density matrix is positive-definite. We then apply this theory to the 1D waveguide-QED system where the squeezing in one direction is experimentally achievable. We show that the enhancement and suppression of the dephasing rate caused by the squeezed vacuum is actually position dependent. In single-atom case, the squeezing does not affect its population dynamics. However, in multi-atom case, the squeezing can strongly affect the population dynamics of the system because two-photon absorption and emission are allowed in multi-atom system. We also show that dipole-dipole interaction influences dephasing rate and we can tune the position of the squeezing source to tune the dephasing rate of the system. Moreover, we show that stationary entangled state can be achieved in this system independent of the initial state and the emitter separation. Particularly, when the center of mass is close to $n\lambda_{0z}/4$ and the squeezing is large, the system can be prepared in GHZ state. Moreover, we study the power spectrum of the resonance fluorescence. It is demonstrated that the phase of the squeezed vacuum, emitter separation, and the center-of-mass position can affect the bandwidth and the intensity of the sidebands.

\section{Acknowledgment}
This work is supported by a grant from the Qatar National Research Fund (QNRF) under NPRP project 8-352-1-074.
\begin{widetext}

\appendix

\section{DERIVATION OF EQ.(6)}
Here we show how to derive the master equation Eq.\eqref{eq6}. We start from a more general case where atoms are not identical but $\omega_{i}\approx \omega_{j}$, and we make the squeezing center frequency $\omega_{0}=\underset{i}{\sum}\omega_{i}/l$. Then we can rewrite the interaction Hamiltonian in Eq.\eqref{eq3} as
\begin{equation}
\label{eqa0}\tag{A1}
V(t)=-i\hbar \sum_{\vec{k}s}[D(t)a_{\vec{k}s}(t)-D^{+}(t)a^{\dagger}_{\vec{k}s}(t)],
\end{equation}
where
\begin{equation}
\label{eqa1}\tag{A2}
\begin{gathered}
 D(t)=\underset{i}{\sum}[\vec{\mu}_{i}\cdot\vec{u}_{\vec{k},s}(r_{i})S_{i}^{\dagger}(t)+\vec{\mu}^{*}_{i}\cdot\vec{u}_{\vec{k},s}(r_{i})S_{i}^{-}(t)]. 
 \end{gathered}
\end{equation}

Since $\left\langle a_{\vec{k},s}\right\rangle =\left\langle a_{\vec{k},s}^{\dagger}\right\rangle =0$, the first term in Eq.\eqref{eq4} vanishes. Therefore, we have
\begin{equation}
\label{eqa2}\tag{A3}
\begin{split}
\frac{d\rho^{S}}{dt}=&-\frac{1}{\hbar^{2}}\int_{0}^{t}d\tau Tr_{F}\{[V(t),[V(t-\tau),\rho^{S}(t-\tau)\rho^{F}\}\\
=&-\frac{1}{\hbar^{2}}\int_{0}^{t}d\tau Tr_{F}\{V(t)V(t-\tau)\rho^{S}(t-\tau)\rho^{F}+\rho^{S}(t-\tau)\rho^{F}V(t-\tau)V(t)\\
&-V(t)\rho^{S}(t-\tau)\rho^{F}V(t-\tau)-V(t-\tau)\rho^{S}(t-\tau)\rho^{F}V(t)\}.
\end{split}
\end{equation} 

Here we just show how to deal with the first term in Eq.\eqref{eqa2}, the remaining terms can be calculated in the same way. For the first term, we have
\begin{equation}
\label{eqa3}\tag{A4}
\begin{split}
&-\frac{1}{\hbar^{2}}\int_{0}^{t}d\tau Tr_{F}\{V(t)V(t-\tau)\rho^{S}(t-\tau)\rho^{F}\}\\
=&\int_{0}^{t}d\tau\underset{\vec{k}s,\vec{k}'s'}{\sum}\{D(t)D(t-\tau)Tr_{F}[\rho^{F}a_{ks}(t)a_{k's'}(t-\tau)]-D(t)D^{+}(t-\tau)Tr_{F}[\rho^{F}a_{ks}(t)a^{\dagger}_{k's'}(t-\tau)]\\
&-D^{+}(t)D(t-\tau)Tr_{F}[\rho^{F}a^{\dagger}_{ks}(t)a_{k's'}(t-\tau)]+D^{+}(t)D^{+}(t-\tau)Tr_{F}[\rho^{F}a^{\dagger}_{ks}(t)a^{\dagger}_{k's'}(t-\tau)]\}\rho^{S}(t-\tau)\}.
\end{split}
\end{equation}
Using Eq.\eqref{eqa1} and the correlation function Eq.\eqref{eq51}$\sim$\eqref{eq55}, under the rotating wave approximation(RWA), we have
\begin{equation}
\label{eqa4}\tag{A5}
\begin{split}
&-\frac{1}{\hbar^{2}}\int_{0}^{t}d\tau Tr_{F}\{V(t)V(t-\tau)\rho^{S}(t-\tau)\rho^{F}\}\\
=& \sum_{ij}\underset{\vec{k}s,\vec{k'}s'}{\sum}\int_{0}^{t}d\tau\{\vec{\mu}{}_{i}\cdot\vec{u}_{\vec{k}s}(r_{i})S_{i}^{+}e^{i\omega_{i}t}\vec{\mu}_{j}\cdot\vec{u}_{\vec{k}'s'}(r_{j})S_{j}^{+}e^{i\omega_{j}(t-\tau)}e^{-i(\omega_{\vec{k}s}+\omega_{\vec{k}'s'})t+i\omega_{\vec{k}'s'}\tau}[-\sinh(r)\cosh(r)\delta_{\vec{k}',2\vec{k}_{0}-\vec{k}}\delta_{ss'}]\\
&-\vec{\mu}_{i}\cdot\vec{u}_{\vec{k}s}(r_{i})S_{i}^{+}e^{i\omega_{i}t}\vec{\mu}^{*}_{j}\cdot\vec{u}_{\vec{k}'s'}^{*}(r_{j})S_{j}^{-}e^{-i\omega_{j}(t-\tau)}e^{-i\omega_{\vec{k}'s'}\tau}\cosh^{2}r\delta_{\vec{k}\vec{k}'}\delta_{ss'}\\
&-\vec{\mu}^{*}_{i}\cdot\vec{u}_{\vec{k}s}(r_{i})S_{i}^{-}e^{-i\omega_{i}t}\vec{\mu}_{j}\cdot\vec{u}^{*}_{\vec{k}'s'}(r_{j})S_{j}^{+}e^{i\omega_{j}(t-\tau)}e^{-i\omega_{\vec{k}'s'}\tau}\cosh^{2}r\delta_{\vec{k}\vec{k}'}\delta_{ss'}\\
&-\vec{\mu}^{*}_{i}\cdot\vec{u}_{\vec{k}s}^{*}(r_{i})S_{i}^{-}e^{-i\omega_{i}t}\vec{\mu}_{j}\cdot\vec{u}_{\vec{k}'s'}(r_{j})S_{j}^{+}e^{i\omega_{j}(t-\tau)}e^{i\omega_{\vec{k}'s'}\tau}\sinh^{2}r\delta_{\vec{k}\vec{k}'}\delta_{ss'}\\
&-\vec{\mu}_{i}\cdot\vec{u}^{*}_{\vec{k}s}(r_{i})S_{i}^{+}e^{i\omega_{i}t}\vec{\mu}^{*}_{j}\cdot\vec{u}_{\vec{k}'s'}(r_{j})S_{j}^{-}e^{-i\omega_{j}(t-\tau)}e^{i\omega_{\vec{k}'s'}\tau}\sinh^{2}r\delta_{\vec{k}\vec{k}'}\delta_{ss'}\\
&+\vec{\mu}^{*}_{i}\cdot\vec{u}_{\vec{k}s}^{*}(r_{i})S_{i}^{-}e^{-i\omega_{i}t}\vec{\mu}^{*}_{j}\cdot\vec{u}^{*}_{\vec{k}'s'}(r_{j})S_{j}^{-}e^{-i\omega_{j}(t-\tau)}e^{i(\omega_{\vec{k}s}+\omega_{\vec{k}'s'})t-i\omega_{\vec{k}'s'}\tau}[-\sinh(r)\cosh(r)\delta_{\vec{k}',2\vec{k}_{0}-\vec{k}}\delta_{ss'}]\}\rho^{S}(t-\tau)
\end{split}
\end{equation}
where we have the relationship $\underset{\vec{k}}{\sum}\rightarrow\frac{L^3}{(2\pi)^3}\int k^{2}dk\int_{\Omega_{k}}$. In Ref. \cite{Scully1997}, it has been shown that
 \begin{equation}
\label{eqa6}\tag{A6}
\begin{split}
\frac{L^{3}}{(2\pi)^{3}}\int k^{2}dk\int_{\Omega_{k}}\underset{s}{\sum}\vec{\mu}{}_{i}\cdot\vec{u}_{\vec{k}s}(r_{i})\vec{\mu}_{j}^{*}\cdot\vec{u}_{\vec{k}s}^{*}(r_{j})\approx\frac{\sqrt{\gamma_{i}\gamma_{j}}}{2\pi\omega_{0}^{3}}\int_{0}^{\infty}d\omega\omega^{3}F(kr_{ij})
 \end{split}
\end{equation}
with
\begin{equation}
\label{eqa7}\tag{A7}
\begin{split}
&F(kr_{ij})=\frac{3}{2}\{[1-\cos^{2}\alpha]\frac{sin(kr_{ij})}{kr_{ij}}+[1-3\cos^{2}\alpha][\frac{\cos(kr_{ij})}{(kr_{ij})^{2}}-\frac{sin(kr_{ij})}{(kr_{ij})^{3}}]\}\\
&\gamma_{i}=\frac{\omega_{i}^{3}\mu_{i}^{2}}{3\pi\epsilon_{0}\hbar c^{3}}
 \end{split}
\end{equation}
where $\vec{r}_{ij}=\vec{r}_i-\vec{r}_j$, $r_{ij}=|\vec{r}_{ij}|$, $\alpha$ is the angle between $\vec{r}_{ij}$ and $\vec{\mu}_{i}$, and the approximation in Eq.\eqref{eqa6} becomes equality when $\omega_{1}=\omega_{2}$.  We can also show that
\begin{equation}
\label{eqa8}\tag{A8}
\begin{gathered}
\frac{L^{3}}{(2\pi)^{3}}\int k^{2}dk\int_{\Omega_{k}}\underset{s}{\sum}\vec{\mu}{}_{i}\cdot\vec{u}_{\vec{k}s}(r_{i})\vec{\mu}_{j}\cdot\vec{u}_{\vec{2k_{0}}-\vec{k},s}(r_{j})\approx\frac{\sqrt{\gamma_{i}\gamma_{j}}}{2\pi\omega_{0}^{3}}\int_{0}^{\infty}d\omega\omega^{2}\sqrt{\omega(2\omega_{0}-\omega)}F(k_{0}|\frac{k}{k_{0}}\vec{r}_{ij}+2\vec{r}_{j}|)e^{2ik_{0}R}
 \end{gathered}
\end{equation}
where R is the distance from the sources to the center mass of two atoms, and the approximation becomes equality when $\omega_{1}=\omega_{2}$. Next, we will show how to calculate the first and the second terms in Eq.\eqref{eqa4}, and the remaining terms can be approached in the same way. Using Eq.\eqref{eqa6}, the second term in Eq.\eqref{eqa4} can be simplified as
\begin{equation}
\label{eqb1}\tag{A10}
\begin{split}
&\underset{\vec{k}s}{\sum}\int_{0}^{t}d\tau\vec{\mu}_{i}\cdot\vec{u}_{\vec{k}s}(r_{i})S_{i}^{+}e^{i\omega_{i}t}\vec{\mu}_{j}^{*}\cdot\vec{u}_{\vec{k}s}^{*}(r_{j})S_{j}^{-}e^{-i\omega_{j}(t-\tau)}e^{-i\omega_{\vec{k}s}\tau}\cosh^{2}r\rho^{S}(t-\tau)\\
&=\cosh^{2}r\frac{\sqrt{\gamma_{i}\gamma_{j}}}{2\pi\omega_{0}^{3}}\int_{0}^{t}d\tau\int_{0}^{\infty}d\omega\omega^{3}F(kr_{ij})e^{i(\omega_{i}-\omega_{j})t}e^{i(\omega_{j}-\omega_{k})\tau}S_{i}^{+}S_{j}^{-}\rho^{S}(t-\tau)
\end{split}
\end{equation}
with $F(kr_{ij})$ given in Eq.\eqref{eqa7}. We here calculate the integral of the first term in $F(kr_{ij})$ ($i\ne j$) and the other terms can be calculated similarly.  
\begin{equation}
\label{eqb2}\tag{A11}
\begin{split}
&\\
&\cosh^{2}r\frac{\sqrt{\gamma_{i}\gamma_{j}} c^{4}}{2\pi\omega_{0}^{3}}\frac{3}{2}\int_{0}^{t}d\tau\int_{0}^{\infty}dkk^{3}\frac{sinkr_{ij}}{kr_{ij}}e^{i(\omega_{j}-\omega_{k})\tau}S_{i}^{+}S_{j}^{-}\rho^{S}(t-\tau)e^{i(\omega_{i}-\omega_{j})t}\\
&=\cosh^{2}r\frac{\sqrt{\gamma_{i}\gamma_{j}}c^{4}}{2\pi\omega_{0}^{3}}\frac{3}{2}\int_{0}^{t}d\tau\int_{-\infty}^{\infty}dkk^{2}\frac{1}{2ir_{ij}}(e^{i(k-k_{j})r_{ij}+ik_{j}r_{ij}}-e^{-i(k-k_{j})r_{ij}-ik_{j}r_{ij}})e^{-i(k-k_{j})c\tau}S_{i}^{+}S_{j}^{-}\rho^{S}(t-\tau)e^{i(\omega_{i}-\omega_{j})t}\\
&\approx \cosh^{2}r\frac{\sqrt{\gamma_{i}\gamma_{j}} c^{4}}{2\pi\omega_{0}^{3}}\frac{3}{2}\int_{0}^{t}d\tau k_{j}^{2}\frac{1}{ir_{ij}}[\delta(r_{ij}-c\tau)e^{ik_{j}r_{ij}}-\delta(r_{ij}+c\tau)e^{-ik_{j}r_{ij}}]S_{i}^{+}S_{j}^{-}\rho^{S}(t-\tau)e^{i(\omega_{i}-\omega_{j})t}\\
&\approx \cosh^{2}r\frac{\sqrt{\gamma_{i}\gamma_{j}} c^{4}}{2\pi\omega_{0}^{3}}\frac{3}{2}k_{j}^{2}\frac{\pi}{icr_{ij}}e^{ik_{j}r_{ij}}S_{i}^{+}S_{j}^{-}\rho^{S}(t)e^{i(\omega_{i}-\omega_{j})t}\\
&\approx \frac{3}{4}\sqrt{\gamma_{i}\gamma_{j}}\cosh^{2}r\frac{e^{ik_{0}r_{ij}}}{ik_{0}r_{ij}}S_{i}^{+}S_{j}^{-}\rho^{S}(t)e^{i(\omega_{i}-\omega_{j})t}\\
\end{split}
\end{equation}
In the second line of the equations, we replace $\int_{0}^{\infty}dk$ by $\int_{-\infty}^{\infty}dk$ since the main contribution comes from the frequency around $\omega_{0}$ and the negative frequency part leads to fast-oscillating term such that its integration $\int_{0}^{t}d\tau$ vanishes. From the second line to the third line, the Weisskopf-Wigner approximation\cite{Scully1997} is applied and $k$ is replaced by $k_j$ because the contribution comes mainly from the resonant frequency. From the third line to the fourth line, we assume that the two atoms are very close that the time-retarded effect can be neglected. In the last line, we use the fact that $\omega_i\approx \omega_0$

The other terms in Eq.\eqref{eqb1} can be calculated in a similar way, and the result is given by 
\begin{equation}
\label{eqb6}\tag{A12}
\begin{split}
\frac{\sqrt{\gamma_{i}\gamma_{j}}}{2\pi\omega_{0}^{3}}\int_{0}^{t}d\tau\int_{0}^{\infty}dkk^{3}F(kr_{ij})e^{i(\omega_{i}-\omega_{j})t}e^{i(\omega_{j}-\omega_{k})\tau}S_{i}^{+}S_{j}^{-}\rho^{S}(t-\tau)=(\frac{1}{2}\gamma_{ij}+i\Lambda_{ij})S_{i}^{+}S_{j}^{-}\rho^{S}(t)e^{i(\omega_{i}-\omega_{j})t}
\end{split}
\end{equation}
where 
\begin{equation}
\label{eqa14}\tag{A13}
\begin{split}
&\Lambda_{ij}=\frac{3}{4}\sqrt{\gamma_{i}\gamma_{j}}\{-(1-\cos^{2}\alpha)\frac{\cos(k_{0}r_{ij})}{k_{0}r_{ij}}+(1-3\cos^{2}\alpha)[\frac{sin(k_{0}r_{ij})}{(k_{0}r_{ij})^{2}}+\frac{\cos(k_{0}r_{ij})}{(k_{0}r_{ij})^{3}}]\}\\
&\gamma_{ij}=\sqrt{\gamma_{i}\gamma_{j}} F(k_{0}r_{ij})\\
\end{split}
\end{equation}
 All the other terms with the combination of $S_{i}^{+}$ and $S_{i}^{-}$ can also be calculated in the same way. Thus, all the thermal terms and oscillation terms in Eq.\eqref{eq6} can be given. 

Next we need to calculate the squeezed vacuum terms including $S_{i}^{+}S_{j}^{+}$ or $S_{i}^{-}S_{j}^{-}$. Here we show the calculation of the first term in Eq.\eqref{eqa4} as an example. By inserting Eq.\eqref{eqa8}, the first term of Eq.\eqref{eqa4} yields 
\begin{equation}
\label{eqb3}\tag{A14}
\begin{split}
&\underset{\vec{k}s,\vec{k'}s'}{\sum}\int_{0}^{t}d\tau\int d^{3}k\{\vec{\mu}{}_{i}\cdot\vec{u}_{2\vec{k}_{0}-\vec{k},s}(r_{i})\vec{\mu}_{j}\cdot\vec{u}_{\vec{k}s}(r_{j})e^{i(\omega_{\vec{k}s}-\omega_{j})\tau}S_{i}^{+}S_{j}^{+}\rho^{S}(t-\tau)\\
&=\frac{\sqrt{\gamma_{i}\gamma_{j}} c^{4}}{2\pi\omega_{0}^{3}}\int_{0}^{t}d\tau\int_{0}^{2k_{0}}dkk^{2}\sqrt{k(2k_{0}-k)}F(k_{0}|\frac{k}{k_{0}}\vec{r}_{ij}+2\vec{r}_{j}|)e^{i(\omega_{k}-\omega_{j})\tau}S_{i}^{+}S_{j}^{+}\rho^{S}(t-\tau)e^{2ik_{0}R}\\
&\approx \frac{\sqrt{\gamma_{i}\gamma_{j}} c}{2\pi}\int_{0}^{t}d\tau\int_{-\infty}^{\infty}dkF(k_{0}|\frac{k}{k_{0}}\vec{r}_{ij}+2\vec{r}_{j}|)e^{i(\omega_{k}-\omega_{0})\tau}S_{i}^{+}S_{j}^{+}\rho^{S}(t-\tau)e^{2ik_{0}R}
\end{split}
\end{equation}
From the second line to the third line, the integral limit is extended to $\pm \infty $ and $k^{2}\sqrt{k(2k_{0}-k)}$ is pulled out as $k_{0}^{3}$ according to the Weisskopf-Wigner approximation. To calculate one term with fixed $i,j$, we need to rebuild the coordinate system where $\vec{r}_{i}+\vec{r}_j=0$ for $i\neq j$(We need to build different coordinate systems for different pairs of $i,j$). For example, we here consider the first two atoms, $i,j=1,2$. When $i=j$, this term directly gives $\frac{1}{2}\gamma \cosh^{2}rF(2k_{0}|\vec{r}_{j}|)S_{i}^{+}S_{i}^{+}\rho^{S}(t)$. When $i\neq j$, since there is a singular point at $k=k_{0}$, the calculation is a little bit more complicated but can still be calculated. We have the following integrals:
\begin{equation}
\label{eqb4}\tag{A15}
\begin{split}
&\int_{-\infty}^{\infty}dk\frac{\sin{kr_{ij}}}{kr_{ij}}e^{-ikc\tau}=\frac{\pi}{r_{ij}}\theta_{1}(r_{ij}-c\tau),\\
&\int_{-\infty}^{\infty}dk \Big [\frac{\cos{kr_{ij}}}{(kr_{ij})^{2}}-\frac{\sin{kr_{ij}}}{(kr_{ij})^{3}}\Big ]e^{-ikc\tau}=\frac{\pi(c\tau-r_{ij})(c\tau+r_{ij})}{2r^{3}_{ij}}\theta_{2}(r_{ij}-c\tau),\\
\end{split}
\end{equation}
where $\theta_{1,2}(x)$ are step functions: $\theta_{1,2}(x)=0$ when $x<0$, $\theta_{1,2}(x)=1$ when $x>0$, and $\theta_{1}(0)=1/2$ and $\theta_{2}(0)=0$.
Since $F(k_{0}|\frac{k}{k_{0}}\vec{r}_{ij}+2\vec{r}_{j}|)=F((k-k_{0})r_{12})$, we have
\begin{equation}
\label{eqb5}\tag{A16}
\begin{split}
&\int_{0}^{t}d\tau\int_{-\infty}^{\infty}dkF([(k-k_0)r_{12}]e^{i(\omega_{0}-\omega_{k})\tau}\rho^{S}(t-\tau) \\ &=\int_{0}^{\frac{r_{ij}}{c}}d\tau\frac{3}{2}[(1-\cos^{2}\alpha)\frac{\pi}{r_{ij}}+(1-3\cos^{2}\alpha)\frac{\pi(c\tau-r_{ij})(c\tau+r_{ij})}{2r_{ij}^{3}}]\rho^{S}(t-\tau) \\ & \approx \frac{\pi}{c}\rho^{S}(t).
\end{split}
\end{equation}
In Eq.\eqref{eqb5}, the emitter separation is assumed to be small and the Markovian approximation is applied such that $\rho^{S}(t-\tau)\approx \rho^{S}(t)$. Hence, Eq.\eqref{eqb3} gives $\sinh{r} \cosh{r}\frac{\gamma'_{ij}}{2} S_{i}^{+}S_{j}^{+}\rho^{S}(t)$ with $\gamma'_{ij}=e^{2ik_{0}R}\gamma F(k_{0}|\vec{r}_{i}+\vec{r}_{j}|)$ after transforming the above results to the original coordinate system(Although replacing $k$ by $k_0$ in Eq.\eqref{eqb3}'s last line yields the same result, it is not always safe to do so since $F(x)$ is an oscillating function). Having Dealt with all the squeezed vacuum terms, we can get 
\begin{equation}
\label{eqb17}\tag{A17}
\begin{split}
\frac{d\rho^{S}}{dt}=&-\frac{1}{2}\sum_{\alpha=\pm}\underset{i,j}{\sum}\gamma'_{ij}M(\rho^{S}S_{i}^{\alpha}S_{j}^{\alpha}+S_{i}^{\alpha}S_{j}^{\alpha}\rho^{S}-2S_{j}^{\alpha}\rho^{S}S_{i}^{\alpha})\\
&-\frac{1}{2}\underset{i,j}{\sum}\gamma{}_{ij}(1+N)(\rho^{S}S_{i}^{+}S_{j}^{-}+S_{i}^{+}S_{j}^{-}\rho^{S}-2S_{j}^{-}\rho^{S}S_{i}^{+})e^{i(\omega_i-\omega_j)t}\\
&-\frac{1}{2}\underset{i,j}{\sum}\gamma{}_{ij}N(\rho^{S}S_{i}^{-}S_{j}^{+}+S_{i}^{-}S_{j}^{+}\rho^{S}-2S_{j}^{+}\rho^{S}S_{i}^{-})e^{-i(\omega_i-\omega_j)t}\\&-i\underset{i\neq j}{\sum}\Lambda_{ij}[S_{i}^{+}S_{j}^{-},\rho^{S}]e^{i(\omega_i-\omega_j)t}
\end{split}
\end{equation}
and Eq.\eqref{eq6} is the special case when $\omega_i=\omega_0$.

\section{POSITIVE DEFINITENESS OF DENSITY MATRIX}
In the following we will show that Eq.\eqref{eq6} can be written in the Lindblad equation and it is positive definite:
\begin{equation}
\label{eqa15}\tag{B1}
\begin{gathered}
\frac{d\rho^{S}}{dt}=-i\underset{i}{\sum}[H,\rho^{S}]+\underset{m,n}{\sum}h_{nm}(L_{n}\rho L_{m}^{\dagger}-\frac{1}{2}(\rho L_{m}^{\dagger}L_{n}+L_{m}^{\dagger}L_{n}\rho))
\end{gathered}
\end{equation}
where 
\begin{equation}
\label{eqa16}\tag{B2}
\begin{split}
&H=\underset{i\neq j}{\sum}\Lambda_{ij}S_{i}^{+}S_{j}^{-}\\
&L_{1}=S_{1}^{+},L_{2}=S_{2}^{+},L_{3}=S_{3}^{-},L_{4}=S_{4}^{-}\\
&h=\begin{bmatrix}
	\gamma_{11}\sinh^{2}{r} &\gamma_{12}\sinh^{2}{r}&\gamma'_{11}\sinh{r}\cosh{r}&\gamma'_{12}\sinh{r}\cosh{r}\\
	\gamma_{12}\sinh^{2}{r}&\gamma_{11}\sinh^{2}{r}&\gamma'_{12}\sinh{r}\cosh{r}&\gamma'_{11}\sinh{r}\cosh{r}\\
	\gamma'_{11}\sinh{r}\cosh{r}&\gamma'_{12}\sinh{r}\cosh{r}&\gamma_{11}\cosh^{2}{r}&\gamma_{12}\cosh^{2}{r}\\
	\gamma'_{12}\sinh{r}\cosh{r}&\gamma'_{11}\sinh{r}\cosh{r}&\gamma_{12}\cosh^{2}{r}&\gamma_{11}\cosh^{2}{r}\\
\end{bmatrix}
\end{split}
\end{equation}
here for simplicity, we have already used the relations: $ \gamma'_{12}=\gamma'_{21}$, $\gamma_{12}=\gamma_{21}$, $\gamma_{11}=\gamma_{22}$, $\gamma'_{11}=\gamma'_{22}$. The last relation $\gamma'_{11}=\gamma'_{22}$ is not always satisfied, but without it we cannot diagonalize matrix $h$ analytically. Hence we set $r_i+r_j=0$. Now matrix $h$ can be diagonalized: 
\begin{equation}
\label{eqa17}\tag{B3}
\begin{split}
&h=u^{\dagger}\begin{bmatrix}
	\zeta_{1}&&&\\
	&\zeta_{2}&&\\
	&&\zeta_{3}&\\
	&&&\zeta_{4}\\
\end{bmatrix}u
\end{split}
\end{equation}
where $u$ is a unitary matrix, and 
\begin{equation}
\label{eqa18}\tag{B4}
\begin{split}
&\zeta_{1}=\frac{1}{2}[(\gamma_{11}-\gamma_{12})(1+2\sinh^{2}r)-\sqrt{(\gamma_{11}-\gamma_{12})^{2}+4\sinh^{2}r\cosh^{2}r(\gamma'_{11}-\gamma'_{12})^{2}}]\\
&\zeta_{2}=\frac{1}{2}[(\gamma_{11}-\gamma_{12})(1+2\sinh^{2}r)+\sqrt{(\gamma_{11}-\gamma_{12})^{2}+4\sinh^{2}r\cosh^{2}r(\gamma'_{11}-\gamma'_{12})^{2}}]\\
&\zeta_{3}=\frac{1}{2}[(\gamma_{11}+\gamma_{12})(1+2\sinh^{2}r)-\sqrt{(\gamma_{11}+\gamma_{12})^{2}+4\sinh^{2}r\cosh^{2}r(\gamma'_{11}+\gamma'_{12})^{2}}]\\
&\zeta_{4}=\frac{1}{2}[(\gamma_{11}+\gamma_{12})(1+2\sinh^{2}r)+\sqrt{(\gamma_{11}+\gamma_{12})^{2}+4\sinh^{2}r\cosh^{2}r(\gamma'_{11}+\gamma'_{12})^{2}}]\\
\end{split}
\end{equation}
We noticed that since $|\gamma_{11}-\gamma_{12}|=|\gamma'_{11}-\gamma'_{12}|$ for $r_i+r_j=0$, none of the eigenvalues is negative, so the density matrix is completely positive for any initial condition. For arbitrary $r_i,r_j$, we can only get the positive eigenvalues numerically. 

\section{DERIVATION OF EQ. (11)}

Now let's consider the perfect rectangular waveguide with cross section $a\times b$. The rectangular waveguide can support both TE and TM electric field modes and they are given as follows(To get a neat expression of field equation, we set the origin of our coordinate system at the corner of the waveguide):
\begin{equation}
\label{eqac1}\tag{C1}
\begin{split}
&E_{z}^{TM}=E_{0}sin\frac{m\pi x}{a}sin\frac{n\pi y}{b}e^{ik_{z}z},\, \, \, \, \, \, \, \, \, \, \, \, \, \, \, \, \, \, \, \, \, \, \, \, \, \, \, \, \, \, \, \, \, \, \, \, H_{z}^{TE}=H_{0}\cos\frac{m\pi x}{a}\cos\frac{n\pi y}{b}e^{ik_{z}z} \\
&E_{x}^{TM}=E_{0}\frac{ik_{z}}{h_{mn}^{2}}\frac{m\pi}{a}\cos\frac{m\pi x}{a}sin\frac{n\pi y}{b}e^{ik_{z}z}, \, \, \, \, \, \,\, \, \, \, \, \,\, \, E_{x}^{TE}=H_{0}\frac{i\omega_{k}\mu}{h_{mn}^{2}}\frac{n\pi}{a}\cos\frac{m\pi x}{a}sin\frac{n\pi y}{b}e^{ik_{z}z}\\
&E_{y}^{TM}=E_{0}\frac{ik_{z}}{h_{mn}^{2}}\frac{n\pi}{a}sin\frac{m\pi x}{a}\cos\frac{n\pi y}{b}e^{ik_{z}z},\, \, \,\, \, \,\, \, \, \, \, \,\, \, \, \, E_{y}^{TE}=-H_{0}\frac{i\omega_{k}\mu}{h_{mn}^{2}}\frac{m\pi}{a}sin\frac{m\pi x}{a}\cos\frac{n\pi y}{b}e^{ik_{z}z} \\
&H_{x}^{TM}=E_{0}\frac{i\omega_{k}\epsilon}{h_{mn}^{2}}\frac{n\pi}{a}sin\frac{m\pi x}{a}\cos\frac{n\pi y}{b}e^{ik_{z}z},\, \, \, \, \, \,\, \, \, \, \, \,\, \, \,H_{x}^{TE}=-H_{0}\frac{ik_{z}}{h_{mn}^{2}}\frac{m\pi}{a}sin\frac{m\pi x}{a}\cos\frac{n\pi y}{b}e^{ik_{z}z} \\
&H_{y}^{TM}=-E_{0}\frac{i\omega_{k}\epsilon}{h_{mn}^{2}}\frac{m\pi}{a}\cos\frac{m\pi x}{a}sin\frac{n\pi y}{b}e^{ik_{z}z},\, \, \,\, \, \,\, \, \,  H_{y}^{TE}=-H_{0}\frac{ik_{z}}{h_{mn}^{2}}\frac{n\pi}{a}\cos\frac{m\pi x}{a}sin\frac{n\pi y}{b}e^{ik_{z}z} 
\end{split}
\end{equation}
where $h_{mn}=\sqrt{(\frac{m\pi}{a})^{2}+(\frac{n\pi}{b})^{2}}$, $\epsilon (\mu )$ is the permittivity (permeability), and $H_{0}, E_{0}$ are arbitrary constants. For quantized modes, we have $E_0=\sqrt{4\hbar h_{mn}^{2}/\epsilon ^{2}\mu \nu LS}$ and $H_0=\sqrt{4\hbar h_{mn}^{2}/\epsilon \mu^{2} \nu LS}$\cite{Kim2013}. The dispersion relation inside the waveguide is given by $\omega_{k}^{2}/c^{2}=(m\pi/a)^{2}+(n\pi/b)^{2}+k_z^{2}$. For simplicity, we here consider the waveguide with square cross section, i.e., $a=b$ and the dispersion curves of different modes are shown in Fig.~\ref{1}(b). For square waveguide, $TE_{mn} (TM_{mn})$ and $TE_{nm} (TM_{nm})$ modes are degenerate, and $TE_{10}$ and $TE_{01}$ have the lowest energy.

We assume that the all emitters' transition frequencies are the same and they are below the cutoff frequency of $TE_{11}$ and $TM_{11}$ modes. Since the rectangular waveguide cannot support the $TM_{10}$ and $TM_{01}$ mode, the emitter can only couple to the $TE_{01}$ or $TE_{10}$ modes. Here, without loss of generality we assume that the transition dipole moment of the emitter is in the $y$ direction. Thus, it can only couple to the $TE_{10}$ mode. The emitters are assumed to be located at the center of the waveguide cross section, i.e., $(\frac{a}{2},\frac{a}{2},r_{i})$ and $(\frac{a}{2},\frac{a}{2},r_{j})$. In this case, the mode function for $TE_{10}$ mode is given by $\vec{u}_{k_{z}}(\vec{r}_{i})=\sqrt{\frac{\omega_{k_{z}}\hbar}{\epsilon_{0}LS}}\hat{y}e^{ik_{z}(\vec{r}-\vec{o}_{k_{z}})}$ with $S=a^2$. By reducing the cross section, we can increase the amplitude of the mode function and therefore the coupling strength.

Compared with the free space case shown in Appendix A, the only modification to the calculation for the waveguide is $\sum_{\vec{k}s}\rightarrow \sum_{k_{z}}$ in Eq.\eqref{eqa4}. We here calculate the first and the second term in Eq.\eqref{eqa4} to show how to get Eq.\eqref{eq6} and Eq.\eqref{eq8}. For the second term, we have
\begin{equation}
\label{eqc2}\tag{C2}
\begin{split}
&-\underset{k_z}{\sum}\int_{0}^{t}d\tau\vec{\mu}_{i}\cdot\vec{u}_{\vec{k}s}(r_{i})S_{i}^{+}e^{i\omega_{0}t}\vec{\mu}_{j}^{*}\cdot\vec{u}_{\vec{k}'s'}^{*}(r_{j})S_{j}^{-}e^{-i\omega_{0}(t-\tau)}e^{-i\omega_{\vec{k}'s'}\tau}\cosh^{2}r\rho^{S}(t-\tau)\delta_{\vec{k}\vec{k}'}\delta_{ss'}\\
=&-\frac{L}{2\pi}\int_{-\infty}^{\infty}dk_{z}\int_{0}^{t}d\tau e^{i\omega_{0}\tau}e^{-i\omega_{k_{z}}\tau}\frac{\omega_{k}\mu^{2}}{\epsilon_{0}LS\hbar}e^{ik_{z}(r_{i}-r_{j})}\cosh^{2}rS_{i}^{+}S_{j}^{-}\rho^{S}(t-\tau)\\
\approx&-\frac{L}{2\pi}\int_{0}^{\infty}dk_{z}\int_{0}^{t}d\tau e^{i\omega_{0}\tau}e^{-i[\omega_{0}+c^{2}k_{0z}(k_{z}-k_{0z})/\omega_{0}]\tau}\frac{\omega_{k}\mu^{2}}{\epsilon_{0}LS\hbar}[e^{ik_{z}(r_{i}-r_{j})}+e^{-ik_{z}(r_{i}-r_{j})}]\cosh^{2}rS_{i}^{+}S_{j}^{-}\rho^{S}(t-\tau)\\
\approx&-\frac{L}{2\pi}\int_{-k_{0z}}^{\infty}d\delta k_{z}\int_{0}^{t}d\tau e^{-i\tau c^{2}k_{0z}\delta k_{z}/\omega_{0}}\frac{\omega_{k}\mu^{2}}{\epsilon_{0}LS\hbar}[e^{i(k_{0z}+\delta k_{z})(r_{i}-r_{j})}+e^{-i(k_{0z}+\delta k_{z})(r_{i}-r_{j})}]\cosh^{2}rS_{i}^{+}S_{j}^{-}\rho^{S}(t-\tau)\\
\approx&-\frac{L}{2\pi}\int_{-\infty}^{\infty}d\delta k_{z}\int_{0}^{t}d\tau e^{-i(c^{2}k_{0z}\delta k_{z}/\omega_{0})\tau}\frac{\omega_{k}\mu^{2}}{\epsilon_{0}LS\hbar}[e^{i(k_{0z}+\delta k_{z})(r_{i}-r_{j})}+e^{-i(k_{0z}+\delta k_{z})(r_{i}-r_{j})}]\cosh^{2}rS_{i}^{+}S_{j}^{-}\rho^{S}(t-\tau)\\
\approx&-\frac{L}{2\pi}\int_{0}^{t}d\tau \frac{\omega_{0}\mu^{2}}{\epsilon_{0}LS\hbar}2\pi[e^{ik_{0z}(r_{i}-r_{j})}\delta((r_{i}-r_{j})-\frac{c^{2}k_{0z}}{\omega_{0}}\tau)+e^{-ik_{0z}(r_{i}-r_{j})}\delta((r_{i}-r_{j})+\frac{c^{2}k_{0z}}{\omega_{0}}\tau)]\cosh^{2}rS_{i}^{+}S_{j}^{-}\rho^{S}(t-\tau)\\
\approx&-\frac{L}{2\pi}e^{ik_{0z}r_{ij}}\frac{\omega_{0}\mu^{2}}{\epsilon_{0}LS\hbar}2\pi\frac{\omega_{0}}{c^{2}k_{0z}}\cosh^{2}rS_{i}^{+}S_{j}^{-}\rho^{S}(t)\\
\approx&-[\frac{\gamma_{1d}}{2}\cos(k_{0z}r_{ij})+i\frac{\gamma_{1d}}{2}sin(k_{0z}r_{ij})]\cosh^{2}rS_{i}^{+}S_{j}^{-}\rho^{S}(t)\\
\equiv &-(\frac{\gamma_{ij}}{2}+i\Lambda_{ij})\cosh^{2}rS_{i}^{+}S_{j}^{-}\rho^{S}(t)
\end{split}
\end{equation}
where emitter separation $r_{ij}=|r_{i}-r_{j}|$, $\gamma_{1d}=2\mu^{2}\omega_{0}^{2}/\hbar\epsilon_{0}Sc^{2}k_{0z}$ is the spontaneous decay rate in the waveguide as is shown in Eq.\eqref{eq7}, $\gamma_{ij}=\gamma_{1d} \cos(k_{0z}r_{ij})$ is the collective decay rate, and $\Lambda_{ij}=\gamma_{1d}\sin(k_{0z}r_{ij})/2$ is the collective energy shift.
In the third line we expand $\omega_{k}=c\sqrt{(\frac{\pi}{a})^{2}+(k_{z})^{2}}$ around $k_{z}=k_{0z}$ since resonant modes provide dominant contributions. In the fifth line we extend the integration $\int_{-k_{0z}}^{\infty}dk_{z}\rightarrow\int_{-\infty}^{\infty}dk_{z}$ because the main contribution comes from the components around $\delta k_{z}=0$. In the next line, Weisskopf-Wigner approximation is used. Thus, we have obtained $\gamma_{ij}$ and $\Lambda_{ij}$ as is shown in Eq.\eqref{eq8}. 

Next we need to calculate the first term (squeezing term) in Eq.\eqref{eqa4}:
\begin{equation}
\label{eqb8}\tag{C4}
\begin{split}
& \underset{k_{z}}{\sum}\int_{0}^{t}d\tau\{\vec{\mu}{}_{i}\cdot\vec{u}_{2\vec{k}_{0}-\vec{k}}(r_{i})S_{i}^{+}\vec{\mu}_{j}\cdot\vec{u}_{\vec{k}}(r_{j})S_{j}^{+}e^{i(\omega_{\vec{k}}-\omega_{0})\tau}[-\sinh(r)\cosh(r)]\rho^{S}(t-\tau) \\
&=-\frac{L}{2\pi}\int_{0}^{2k_{0z}}dk_{z}\int_{0}^{t}d\tau e^{i(\omega_{k_{z}}-\omega_{0})\tau}e^{i(2k_{0z}-k_{z})(r_{i}-o_{1})}e^{ik_{z}(r_{j}-o_{1})}\frac{\sqrt{\omega_{k_{z}}\omega_{2k_{0z}-k_{z}}}\mu^{2}}{\epsilon_{0}LS\hbar}\sinh(r)\cosh(r)S_{i}^{+}S_{j}^{+}\rho^{S}(t-\tau)\\ &-\frac{L}{2\pi}\int_{-2k_{0z}}^{0}dk_{z}\int_{0}^{t}d\tau e^{i(\omega_{k_{z}}-\omega_{0})\tau}e^{i(-2k_{0z}-k_{z})(r_{i}-o_{2})}e^{ik_{z}(r_{j}-o_{2})}\frac{\sqrt{\omega_{k_{z}}\omega_{-2k_{0z}-k_{z}}}\mu^{2}}{\epsilon_{0}LS\hbar}\sinh(r)\cosh(r)S_{i}^{+}S_{j}^{+}\rho^{S}(t-\tau).
\end{split}
\end{equation}
For $i=j$, Eq.\eqref{eqb8} reduces to 
\begin{equation}
\label{eqb9}\tag{C5}
\begin{split}
&\underset{k_{z}}{\sum}\int_{0}^{t}d\tau\{\vec{\mu}{}_{i}\cdot\vec{u}_{2\vec{k}_{0}-\vec{k}}(r_{i})S_{i}^{+}\vec{\mu}_{j}\cdot\vec{u}_{\vec{k}}(r_{j})S_{j}^{+}e^{i(\omega_{\vec{k}}-\omega_{0})\tau}[-\sinh(r)\cosh(r)]\rho^{S}(t-\tau)\\
& =-\frac{L}{2\pi}\int_{0}^{2k_{0z}}dk_{z}\int_{0}^{t}d\tau e^{i\frac{c^{2}k_{0z}}{_{\omega_{0}}}(k_{z}-k_{0z})\tau}e^{i2k_{0z}(r_{i}-o_{1})}\frac{\sqrt{\omega_{k_{z}}\omega_{2k_{0z}-k_{z}}}\mu^{2}}{\epsilon_{0}LS\hbar}\sinh(r)\cosh(r)S_{i}^{+}S_{j}^{+}\rho^{S}(t-\tau)\\
& -\frac{L}{2\pi}\int_{-2k_{0z}}^{0}dk_{z}\int_{0}^{t}d\tau e^{i\frac{c^{2}k_{0z}}{_{\omega_{0}}}(k_{z}-k_{0z})\tau}e^{-i2k_{0z}(r_{i}-o_{2})}\frac{\sqrt{\omega_{k_{z}}\omega_{-2k_{0z}-k_{z}}}\mu^{2}}{\epsilon_{0}LS\hbar}\sinh(r)\cosh(r)S_{i}^{+}S_{j}^{+}\rho^{S}(t-\tau)\\
&=-\frac{L}{2\pi}[e^{i2k_{0z}(r_{i}-o_{1})}+e^{-i2k_{0z}(r_{i}-o_{2})}]\frac{\omega_{k_{0z}}\mu^{2}}{\epsilon_{0}LS\hbar}\int_{0}^{t}d\tau2\pi\delta(\frac{c^{2}k_{0z}}{\omega_{0}}\tau)\sinh(r)\cosh(r)S_{i}^{+}S_{j}^{+}\rho^{S}(t-\tau)\\
&=-\frac{L}{2\pi}[e^{i2k_{0z}(r_{i}-o_{1})}+e^{-i2k_{0z}(r_{i}-o_{2})}]\frac{\omega_{k_{0z}}\mu^{2}}{\epsilon_{0}LS\hbar}\int_{0}^{t}d\tau2\pi\delta(\frac{c^{2}k_{0z}}{\omega_{0}}\tau)\sinh(r)\cosh(r)S_{i}^{+}S_{j}^{+}\rho^{S}(t-\tau)\\
&=-e^{i2k_{0z}R}\frac{\omega_{0}^{2}\mu^{2}}{\epsilon_{0}\hbar Sc^{2}k_{0z}}\cos(2k_{0z}r_{i})\sinh(r)\cosh(r)S_{i}^{+}S_{j}^{+}\rho^{S}(t)\\
&=-e^{i2k_{0z}R}\frac{\gamma_{1d}}{2}\cos(2k_{0z}r_{i})\sinh(r)\cosh(r)S_{i}^{+}S_{j}^{+}\rho^{S}(t)
\end{split}
\end{equation}
where we have used the fact that the origin of coordinate system is at equal distant from two sources(i.e., $o_2=-o_1=R$) in the second last line. Thus, we have $\gamma'_{ii}=\gamma_{1d}\cos(2k_{0z}r_{i})$. For $i\neq j$, Eq. \eqref{eqb8} reduces to
\begin{equation}
\label{eqb10}\tag{C6}
\begin{split}
&\underset{k_{z}}{\sum}\int_{0}^{t}d\tau\{\vec{\mu}{}_{i}\cdot\vec{u}_{2\vec{k}_{0}-\vec{k}}(r_{i})S_{i}^{+}\vec{\mu}_{j}\cdot\vec{u}_{\vec{k}}(r_{j})S_{j}^{+}e^{i(\omega_{\vec{k}}-\omega_{0})\tau}[-\sinh(r)\cosh(r)]\rho^{S}(t-\tau)\\
& =-\frac{L}{2\pi}\int_{0}^{2k_{0z}}dk_{z}\int_{0}^{t}d\tau e^{i\frac{c^{2}k_{0z}}{_{\omega_{0}}}(k_{z}-k_{0z})\tau}e^{i2k_{0z}(r_{c}-o_{1})}e^{-i(k_{z}-k_{0z})(r_{i}-r_{j})}\frac{\sqrt{\omega_{k_{z}}\omega_{2k_{0z}-k_{z}}}\mu^{2}}{\epsilon_{0}LS\hbar}\sinh(r)\cosh(r)S_{i}^{+}S_{j}^{+}\rho^{S}(t-\tau) \\
& -\frac{L}{2\pi}\int_{-2k_{0z}}^{0}dk_{z}\int_{0}^{t}d\tau e^{i\frac{c^{2}k_{0z}}{_{\omega_{0}}}(-k_{z}-k_{0z})\tau}e^{-i2k_{0z}(r_{c}-o_{2})}e^{-i(k_{z}+k_{0z})(r_{i}-r_{j})}\frac{\sqrt{\omega_{k_{z}}\omega_{-2k_{0z}-k_{z}}}\mu^{2}}{\epsilon_{0}LS\hbar}\sinh(r)\cosh(r)S_{i}^{+}S_{j}^{+}\rho^{S}(t-\tau)\\
& =-\frac{L}{2\pi}\int_{0}^{2k_{0z}}dk_{z}\int_{0}^{t}d\tau e^{i\frac{c^{2}k_{0z}}{_{\omega_{0}}}(k_{z}-k_{0z})\tau}e^{i2k_{0z}(r_{c}-o_{1})}e^{-i(k_{z}-k_{0z})(r_{i}-r_{j})}\frac{\sqrt{\omega_{k_{z}}\omega_{2k_{0z}-k_{z}}}\mu^{2}}{\epsilon_{0}LS\hbar}\sinh(r)\cosh(r)S_{i}^{+}S_{j}^{+}\rho^{S}(t-\tau) \\
& -\frac{L}{2\pi}\int_{0}^{2k_{0z}}dk_{z}\int_{0}^{t}d\tau e^{i\frac{c^{2}k_{0z}}{_{\omega_{0}}}(k_{z}-k_{0z})\tau}e^{-i2k_{0z}(r_{c}-o_{2})}e^{-i(-k_{z}+k_{0z})(r_{i}-r_{j})}\frac{\sqrt{\omega_{-k_{z}}\omega_{-2k_{0z}+k_{z}}}\mu^{2}}{\epsilon_{0}LS\hbar}\sinh(r)\cosh(r)S_{i}^{+}S_{j}^{+}\rho^{S}(t-\tau) \\
& =-\frac{L}{2\pi}e^{i2k_{0z}(r_{c}-o_{1})}\frac{\omega_{k_{0z}}\mu^{2}}{\epsilon_{0}LS\hbar}\int_{-\infty}^{\infty}dk_{z}\int_{0}^{t}d\tau e^{i\frac{c^{2}k_{0z}}{_{\omega_{0}}}(k_{z}-k_{0z})\tau}e^{-i(k_{z}-k_{0z})(r_{i}-r_{j})}\sinh(r)\cosh(r)S_{i}^{+}S_{j}^{+}\rho^{S}(t-\tau) \\
&-\frac{L}{2\pi}e^{-i2k_{0z}(r_{c}-o_{2})}\frac{\omega_{k_{0z}}\mu^{2}}{\epsilon_{0}LS\hbar}\int_{-\infty}^{\infty}dk_{z}\int_{0}^{t}d\tau e^{i\frac{c^{2}k_{0z}}{_{\omega_{0}}}(k_{z}-k_{0z})\tau}e^{i(k_{z}-k_{0z})(r_{i}-r_{j})}\sinh(r)\cosh(r)S_{i}^{+}S_{j}^{+}\rho^{S}(t-\tau)\\
&=-\frac{L}{2\pi}e^{i2k_{0z}R}\frac{\omega_{0}\mu^{2}}{\epsilon_{0}LS\hbar}\int_{0}^{t}d\tau2\pi[e^{i2k_{0z}r_{c}}\delta(r_{i}-r_{j}-\frac{c^{2}k_{0z}}{_{\omega_{0}}}\tau)+e^{-i2k_{0z}r_{c}}\delta(r_{i}-r_{j}+\frac{c^{2}k_{0z}}{_{\omega_{0}}}\tau)]\sinh(r)\cosh(r)S_{i}^{+}S_{j}^{+}\rho^{S}(t-\tau) \\
&=-e^{i2k_{0z}R}\frac{\omega_{0}^{2}\mu^{2}}{\epsilon_{0}\hbar Sc^{2}k_{0z}}e^{i2k_{0z}r_{c}sgn(i-j)}S_{i}^{+}S_{j}^{+}\rho^{S}(t)\rightarrow-\frac{\gamma_{1d}}{2}e^{i2k_{0z}R}\cos(k_{0z}(r_{i}+r_{j}))S_{i}^{+}S_{j}^{+}\rho^{S}(t)
\end{split}
\end{equation}
where $sgn(i-j)$ is the sign function. The last arrow is because we need to sum over $i,j$, so the imaginary part of $e^{i2k_{0z}r_{c}sgn(i-j)}$ vanishes and the neat result is that $\gamma'_{ij}=e^{i2k_{0z}R}\gamma_{1d}\cos(k_{0z}(r_{i}+r_{j}))$. As for $S_{i}^{+}\rho^{S}(t)S_{j}^{+}$ terms, the combination of the last two terms in Eq.\eqref{eqa2} will make the imaginary part of $e^{i2k_{0z}r_{c}sgn(i-j)}$ vanish. Thus, we have $\gamma'_{ij}=e^{i2k_{0z}R}\gamma_{1d}\cos(k_{0z}(r_i+r_j))$. If one needs to get $\gamma_{ij}, \gamma'_{ij}$ and $\Lambda_{ij}$in the unidirectional waveguide case, we just need to discard the second terms in the parenthesis of Eq.\eqref{eqc2} and Eq.\eqref{eqb10}

\end{widetext}

\end{document}